\documentclass{article}

\PassOptionsToPackage{numbers, compress}{natbib}

\usepackage[final]{neurips_2023}

\usepackage{multirow}
\usepackage{booktabs}

\usepackage[utf8]{inputenc} % allow utf-8 input
\usepackage[T1]{fontenc}    % use 8-bit T1 fonts
\usepackage{hyperref}       % hyperlinks
\usepackage{url}            % simple URL typesetting
\usepackage{booktabs}       % professional-quality tables
\usepackage{amsfonts}       % blackboard math symbols
\usepackage{nicefrac}       % compact symbols for 1/2, etc.
\usepackage{microtype}      % microtypography
\usepackage{xcolor}         % colors
\usepackage{amsmath}
\usepackage{graphicx} 

\title{\textit{LANE}: \textit{L}ogic \textit{A}lignment of \textit{N}on-tuning Large Language Models and Online Recommendation Systems for \textit{E}xplainable Reason Generation}

\author{%
  Hongke Zhao \thanks{These authors contributed equally to this work.} \\
  Tianjing University \\
  \And
  Songming Zheng \footnotemark[1] \\
  Tianjing University \\
  \AND
  Likang Wu \thanks{Corresponding author.}\\
  Tianjing University \\
  \And
  Bowen Yu \\
  Baidu Talent Intelligence Center, Baidu Inc \\
  \texttt{yubowen04@baidu.com} \\
  \And
  Jing Wang \\
  Baidu Talent Intelligence Center, Baidu Inc \\
  \texttt{wangjing79@baidu.com} \\
}

\begin{document}

\maketitle

\begin{abstract}
The explainability of recommendation systems is crucial for enhancing user trust and satisfaction. Leveraging large language models (LLMs) offers new opportunities for comprehensive recommendation logic generation. However, in existing related studies, fine-tuning LLM models for recommendation tasks incurs high computational costs and alignment issues with existing systems, limiting the application potential of proven proprietary/closed-source LLM models, such as GPT-4. In this work, our proposed effective strategy \textbf{LANE} aligns LLMs with online recommendation systems without additional LLMs tuning, reducing costs and improving explainability. This innovative approach addresses key challenges in integrating language models with recommendation systems while fully utilizing the capabilities of powerful proprietary models. Specifically, our strategy operates through several key components: semantic embedding, user multi-preference extraction using zero-shot prompting, semantic alignment, and explainable recommendation generation using Chain of Thought (CoT) prompting. By embedding item titles instead of IDs and utilizing multi-head attention mechanisms, our approach aligns the semantic features of user preferences with those of candidate items, ensuring coherent and user-aligned recommendations. Sufficient experimental results including performance comparison, questionnaire voting, and visualization cases prove that our method can not only ensure recommendation performance, but also provide easy-to-understand and reasonable recommendation logic.

\end{abstract}

\section{Introduction}
\label{1}

In the realm of recommendation systems, the explainability of results has emerged as a critical factor. The importance of explainability lies in its ability to enhance user trust and satisfaction by providing clear and understandable reasons behind recommendations \cite{tintarev2015explaining,gedikli2014should,zhang2020explainable}. The advent of large language models (LLMs)  with their robust language comprehension and generation capabilities opens new avenues for bolstering the explainability of recommendation systems\cite{hou2024large,wei2024llmrec,guan2024enhancing}.

Utilizing large language models to generate recommendation reasons involves fine-tuning or prompt-tuning approaches \cite{fan2023recommender,guan2024langtopo,liu2024dr}. Model tuning allows LLMs to adjust their parameters based on specific datasets , aligning their outputs closely with the desired recommendation logic \cite{wu2024exploring,cheng2023explainable,bao2023tallrec,lin2024data}. Through this direct way, LLMs can effectively summarize and learn user preferences inherent in recommendation tasks. Consequently, LLMs can function both as recommenders and explainers, or serve as auxiliary explainers to augment the explainability of existing online recommendation systems, such as SASrec \cite{kang2018self}.

However, these tuning strategies are not without their drawbacks. Firstly, compared to traditional recommendation models, training and updating LLMs incur substantial computational resource costs \cite{strubell2019energy}. The significant computational demands translate to higher energy consumption and increased operational expenses, posing practical challenges for widespread deployment. Moreover, representative proprietary models, like ChatGPT-4 , exhibit superior language generation and commercial application capabilities relative to open-source models \cite{liu2023summary}. Nevertheless, there is a misalignment between the user preferences inferred through historical sequence prompts and those used by online recommendation systems, leading to inconsistent recommendation logic. This discrepancy implies that proprietary models cannot be effectively utilized in practical explainable recommendation tasks if they require tuning for alignment. 

To address these issues, our objective is to propose a novel learning strategy that aligns the recommendation logic of LLMs with that of online recommendation systems without necessitating the training of the LLMs themselves. This approach aims to significantly reduce model training and maintenance costs while harnessing the potential of proprietary commercial models to enhance the explainability of existing recommendation systems. Achieving this goal represents a significant breakthrough in combining large models with recommendation systems, overcoming a critical application bottleneck.

In this paper, we propose an innovative explainable recommendation framework \textbf{LANE} that leverages large language models~\cite{brown2020language,devlin2018bert,touvron2023llama} requiring no tuning to align with the recommendation logic of ordinary recommenders. This framework generates reasonable and easily understandable explanations of recommendation reasons. Our primary strategy involves using large language models to sample potential user preferences from multiple perspectives. These preferences are then matched with the embeddings of the operational recommendation system using a query-based learning approach. This straightforward and effective method ensures that the textual explanations of user preferences are consistent with the recommendation logic of the actual system. Our framework is not only model-agnostic but also offers good model explainability. Additionally, to simplify the use of advanced LLMs and fully utilize their capabilities, avoiding issues such as closed-source LLMs or limited computational resources, this LLM part does not require tuning. Instead, it directly employs pre-trained LLMs (such as GPT-4 \cite{achiam2023gpt}) to generate explanation information. The primary technology improvements of different key components in our framework are as follows:

\begin{itemize}
    \item \textbf{Design of the Explainable Framework}: We design a model-agnostic and efficient explainable recommendation framework that uses large language models to generate highly explainable personalized recommendation statements.
    % \item \textbf{Semantic Embedding of Items}: To enhance model explainability and ensure that the user preferences and items extracted by LLMs are embedded in the same semantic space, our proposed framework is title-based. Instead of using IDs, items are input using their titles.
    \item \textbf{Capturing Users' Multi-Preferences}: We utilize the excellent in-context learning (ICL) ability of LLMs and have meticulously designed a zero-shot prompt template for extracting users' multi-preferences. This template guides LLMs to capture the Multi-Preferences inherent in the users' historical interaction sequences without providing any examples.
    \item \textbf{Preference Semantic Alignment}: We propose an attention-based learning strategy to align the reasoning logic of LLM explainer and recommender, which automatically selects the consistent and reasonable user's preference.
    \item \textbf{Generation of Personalized Recommendation Texts}: To ensure that the generated recommendation results have clearer explainability, we designed a Chain of Thought (CoT) prompt template. This template enhances the reasoning process of LLMs to improve the quality of text generation. It guides LLMs step-by-step, analyzing the origins of multiple user preferences, the characteristics of recommended items, their alignment with user preferences, and the generation of personalized recommendation texts.
\end{itemize}

\section{Related Work}
We introduced related representative studies concisely in this section.
\subsection{Explainable Recommendation}
In the field of recommendation systems, explainable recommendation \cite{zhang2020explainable} has become an important research direction. Explainable recommendation systems refer to systems that not only provide personalized recommendations but also explain the reasons and logic behind the recommendations, enabling users to understand how recommendations are made.

In general, explainable recommendation systems could be divided into two major categories. The first category is embedded explainable recommendation systems, which can be further subdivided into several subclasses based on the methods they employ, including factorization \cite{zhang2014explicit,tao2019fact,cheng2019incorporating,chen2018attention}, topic modeling \cite{mcauley2013hidden,ren2017social}, graph \cite{he2015trirank,heckel2017scalable,wang2018tem,wu2023learning}, knowledge graph \cite{xian2019reinforcement,ma2019jointly}, and other deep learning ways \cite{chen2019personalized,balog2019transparent}. The other category is post-hoc explainable recommendation systems, which primarily rely on rules, retrieval, or generation models to generate explanations. For example, Peake et al. \cite{peake2018explanation} proposed an association rule mining method to implement post-hoc explainable recommendations. Singh et al. \cite{singh2018posthoc} investigated post-hoc explanations using a learning-to-rank algorithm based on web search. Wang et al. \cite{wang2018reinforcement} introduced a model-agnostic reinforcement learning framework that can generate sentence explanations for any recommendation model.

\subsection{Prompting LLMs For Recommendation}
Current research utilizing LLMs for recommendations can be broadly categorized into three paradigms \cite{fan2023recommender,wu2023survey}: pre-training, fine-tuning, and prompting. The prompting paradigm, being the most recent, adapts LLMs to specific downstream tasks (such as Top-N recommendation and explainable recommendation) through prompts. This paradigm includes three representative methods: in-context learning, prompt tuning \cite{shin2020autoprompt,li2023personalized}, and instruction tuning \cite{bao2023tallrec, wu2024exploring}. Our research falls under the in-context learning method within the prompting paradigm, which allows LLMs to perform recommendation tasks without any fine-tuning. 

Increasingly, advanced techniques like In-context Learning (ICL) \cite{brown2020language,radford2019language,dong2022survey} and Chain of Thought (CoT) \cite{wei2022chain,kojima2022large} are being explored to manually design prompts for various recommendation tasks. For example, He et al. \cite{he2023large} proposed leveraging LLMs as zero-shot conversational recommender systems (CRS) and introduced numerous exploratory tasks to investigate the mechanisms underlying the remarkable performance of LLMs in conversational recommendations. Liu et al. \cite{liu2023chatgpt} evaluated the performance of ChatGPT on various recommendation tasks by performing ICL on corresponding input-output examples for each task, without fine-tuning. InteRecAgent \cite{huang2023recommender} and RecMind \cite{wang2023recmind} employing CoT prompts, enable LLMs to act as agents, handling complex recommendation tasks.

\section{Methodology}
\label{headings}

\subsection{Problem Statement}
This paper primarily focuses on the prevalent task of sequential recommendation and conducts research based on this task. In the setup of our explainable sequential recommendation problem, we assume a user set \( U=\{u_1, u_2, . . . , u_{|U|} \} \) and an item set \( I=\{i_1, i_2, . . . , i_{|I|} \} \), where \( |U| \) and \( |I| \) represent the number of users and items, respectively. Given the historical interaction sequence \( S^u= \{S_1^u, S_1^u, . . . , S_{|S^u|}^u\} \) for user \( u\in U \), where \( S_t^u\in I \) denotes the interacted item by user \( u \) at time step \( t \), and \( |S^u| \) denotes the length of the sequence. We aim to take the sequence \( \{S_1^u, S_1^u, . . . , S_{(|S^u| - 1)}^u\} \) as input to predict the next interaction item \( S_{|S^u|}^u \) for user \( u \), and generate personalized explainable reason \( \text{Explanation}^u \) about \( S_{|S^u|}^u \) to explain the prediction results.

\subsection{LANE}
\subsubsection{Overview}

\begin{figure}[ht]
    \centering
    \includegraphics[width=1\textwidth]{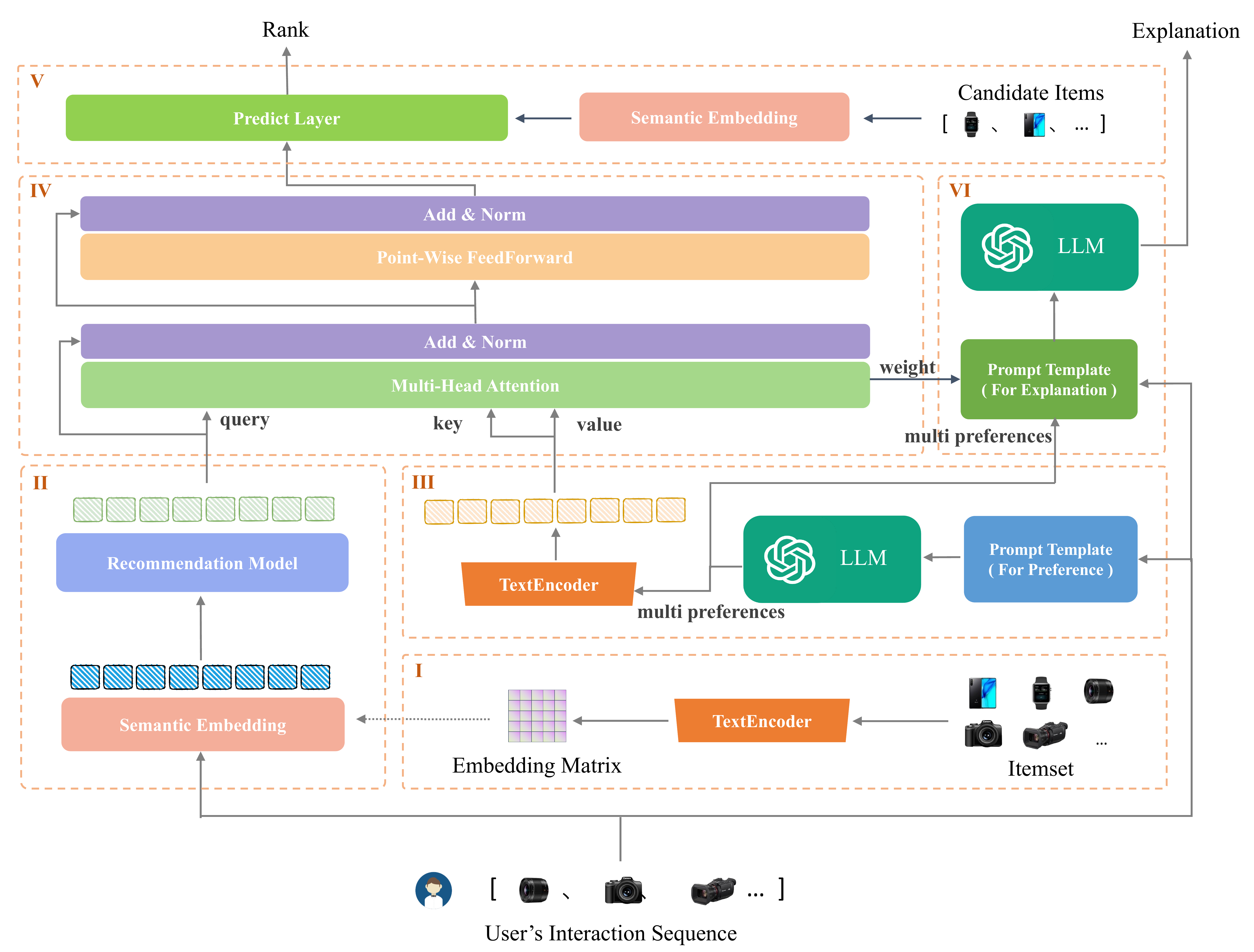}
    \caption{The overview of LANE. It consists of six crucial components: (I) semantic embedding module, (II) integrated model module, (III) users' multi-preference generation module, (IV) semantic alignment module, (V) prediction module, and (VI) explainable recommendation generation module.}
    \label{fig:framework}
\end{figure}

Our proposed explainable recommendation framework LANE is shown in Figure \ref{fig:framework}. The model acquires semantic embeddings of item titles using a text encoder and utilizes these embeddings to initialize the embedding layer of the integrated recommendation model. User interaction sequences are input into predefined prompt templates to guide the LLM in extracting Multi-Preferences, which are then semantically embedded using the same text encoder. A multi-head attention mechanism aligns the semantic features of user sequences with their Multi-Preferences. The aligned vectors and candidate item embeddings are then used to compute recommendation scores, resulting in the final item ranking. Additionally, the model generates explanation information for the recommendations by inputting the user interaction sequences, the previously obtained attention weights, and the multiple preferences into predefined prompt templates.
\subsubsection{Semantic Embedding Module}
\label{3.2.2}
Our proposed framework relies on the powerful language understanding and generation capabilities of large language model (LLM) for the explanation generation part of the recommendation results. To ensure that the users' historical interaction sequence contains more semantic information and to facilitate the understanding of this sequence by the large model, unlike conventional ID-based recommendation models, our proposed model is text(title)-based. Hence, the user interaction sequence \( S^u \) we use is no longer a sequence sorted by item IDs but by item titles. As \( S^u \) is a text sequence, we need to encode it while preserving its semantic information. Given the excellent performance of Sentence-bert in sentence embedding \cite{reimers-2019-sentence-bert}, we choose it as the TextEncoder. By inputting all item titles into the TextEncoder, we capture an embedding matrix:

\begin{equation}
\mathbf{M} = \left[ \begin{array}{c} \mathbf{m}_{1} \\ \mathbf{m}_{2} \\ \vdots \\ \mathbf{m}_{n} \end{array} \right] = \text{TextEncoder} \left( \left[ \begin{array}{c} i_{1} \\ i_{2} \\ \vdots \\ i_{n} \end{array} \right] \right),
\end{equation}

where \( M \in \mathbb{R}^{|I| \times d} \) denotes the embedding matrix, \( d \) denotes the embedding dimension, \( \mathbf{m}_k \in \mathbb{R}^d \) denotes the embedding vector of the \( k \)-th item title, and \( \mathbf{i}_k \) represents the title of the \( k \)-th item. \( \text{TextEncoder}( \cdot ) \) refers to the Sentence-bert model. By retrieving the embedding matrix \( M \), and applying truncation or padding, we can transform the user interaction sequence \( \{S_1^u, S_2^u, . . . , S_{(|S^u| - 1)}^u\} \) into a fixed-length vector sequence \( \mathbf{s}^u = \{ \mathbf{s}_1^u, \mathbf{s}_2^u, . . . , \mathbf{s}_n^u \} \), where \( \mathbf{s}_t^u \in \mathbb{R}^d \) denotes the embedding vector of item \( S_t^u \), \( n \) denotes the fixed-length of sequence. If the sequence length exceeds \( n \), we extract the last \( n \) items from the sequence. Conversely, if the sequence length is less than \( n \), we prepend a padding zero vector \( \mathbf{0} \) until the sequence reaches the desired length \( n \).

\subsubsection{Integrated Model Module}
The primary objective of our proposed framework is to leverage large-scale models to achieve explainability for recommendations generated by traditional black-box models. 
Therefore, our framework needs to integrate sequential recommendation models as the objects to be explained. Among non-explainable sequential recommendation models, SASRec has demonstrated outstanding performance across numerous sequential recommendation datasets and applications \cite{kang2018self}, making it highly representative. \textbf{Hence, we have selected SASRec as an example to illustrate this module, and other recommendation models follow a similar development.}

Recommender is the target we aim to empower, we largely retain all settings of SASRec and only make adaptive modifications to the following two parts:
1). Since SASRec is an ID-based model, we have made modifications to its embedding layer. The embedding layer of SASRec is no longer initialized randomly but with the previously obtained embedding matrix \( M \) to preserve the original semantic information. The position encoding of the embedding layer is still retained, and it is a randomly initialized learnable embedding matrix.
2). The recommender serves as an embedding part of our framework, it does not need to output the final prediction results (the ranking scores for each candidate item). Instead, SASRec only needs to output the feature vectors of the user sequences used for calculating the ranking scores. Therefore, we have also modified the prediction layer to directly return the feature vectors of the sequences.

We maintain consistency with the original implementation of SASRec for other aspects, and specific implementation details can be referred to \cite{kang2018self}. Given the interaction sequence \( \{ \mathbf{s}_1^u, \mathbf{s}_2^u, . . . , \mathbf{s}_n^u \} \), where \( \mathbf{s}_t^u \in \mathbb{R}^d \), the formulation can be expressed as follows:

\begin{equation}
\mathbf{E}^u = \left[ \begin{array}{c} \mathbf{e}_{1}^u \\ \mathbf{e}_{2}^u \\ \vdots \\ \mathbf{e}_{n}^u \end{array} \right] = \left[ \begin{array}{c} \mathbf{s}_{1}^u + \mathbf{PE}_{1} \\ \mathbf{s}_{2}^u + \mathbf{PE}_{2} \\ \vdots \\ \mathbf{s}_{n}^u + \mathbf{PE}_{n} \end{array} \right],
\end{equation}

\begin{equation}
\mathbf{Q}^u = \left[ \begin{array}{c} \mathbf{q}_{1}^u \\ \mathbf{q}_{2}^u \\ \vdots \\ \mathbf{q}_{n}^u \end{array} \right] = \text{SASRec}_{\text{Adapted}}(\mathbf{E}^u) ,
\end{equation}

where \( \mathbf{PE}_t \in \mathbb{R}^d \) denotes the positional encoding at time step \( t \) in the sequence, \( \mathbf{E}^u \in \mathbb{R}^{n \times d} \) denotes the input embedding matrix of the interaction sequence for user \( u \), \( \text{SASRec}_{\text{Adapted}}(\cdot) \) denotes the SASRec model adapted after modifications, \( \mathbf{Q}^u \in \mathbb{R}^{n \times d} \) denotes the feature matrix of the interaction sequence for user \( u \), and \( \mathbf{q}_t^u \in \mathbb{R}^d \) denotes the feature vector of the subsequence consisting of the first \( t \) items for user \( u \).

\subsubsection{Users' Multi-Preferences Generation Module}
\begin{figure}[t]
    \centering
    \includegraphics[width=0.95\textwidth]{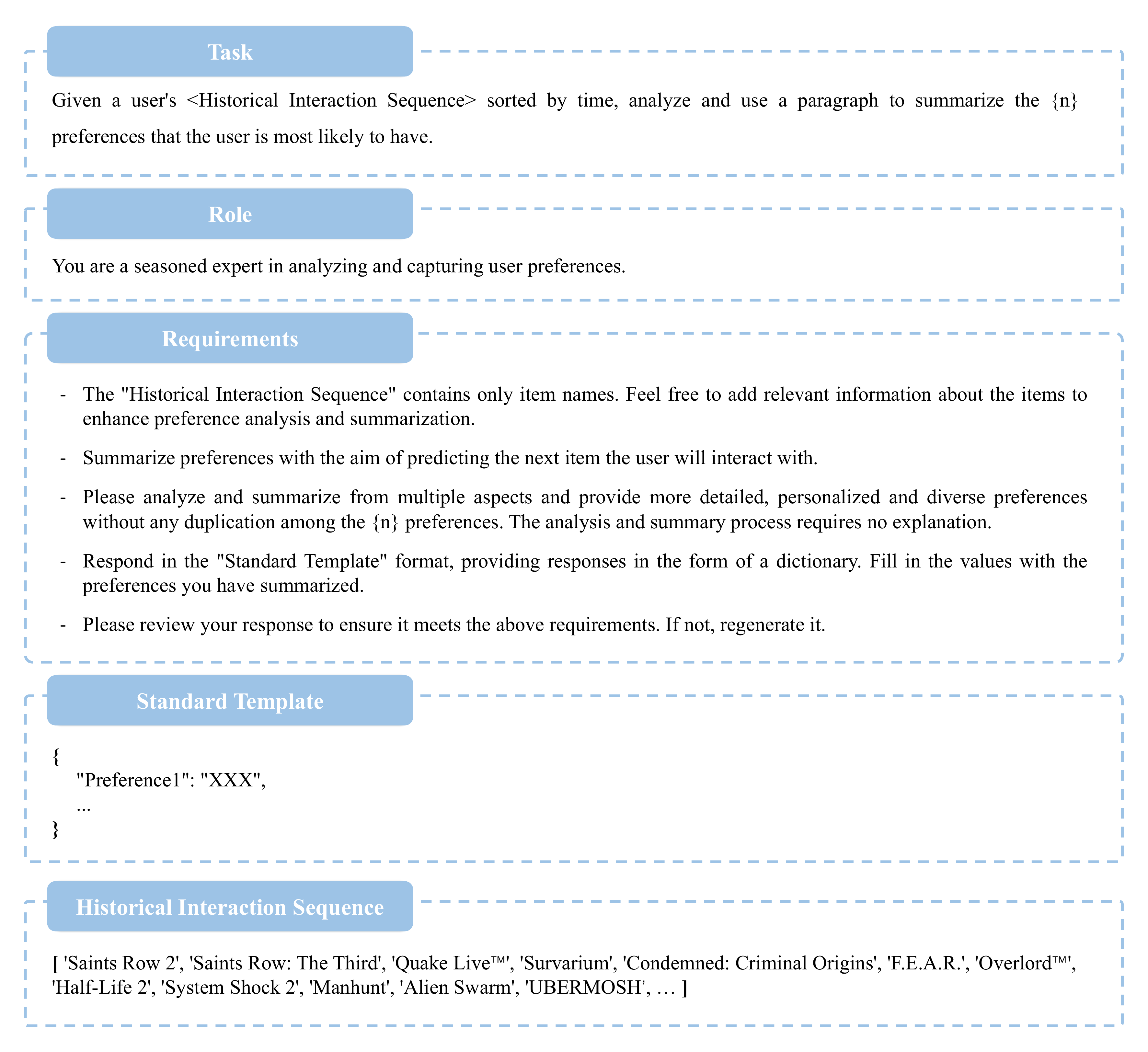}
    \caption{The zero-shot prompt template. It consists of five components and fills in a user interaction sequence in the Steam dataset as an example, where \(n\) refers to the number of user preferences.}
    \label{fig:zero_shot_prompt}
\end{figure}

In recent years, significant breakthroughs have been achieved in natural language model research, leading to the emergence of many large-scale models with outstanding language understanding and generation capabilities. These models have been widely applied across various domains, including recommendation systems. GPT is a notable representative in this regard. Given LLM's remarkable In-Context Learning (ICL) ability \cite{brown2020language}, we leverage zero-shot prompting to guide LLM in extracting features from user interaction sequences and generating human-understandable feature texts—Multi-Preferences. To achieve this, we design a zero-shot prompt template to capture the Multi-Preferences contained in user historical interaction sequences, as illustrated in Figure \ref{fig:zero_shot_prompt}.

The prompt template is composed of five components: Task, Role, Requirements, Standard Template, and Historical Interaction Sequence. In the Task component, we briefly describe the task we need the large model to accomplish. In the Role component, we specify the role of the large model as a ``seasoned expert in analyzing and capturing user preferences'' to enhance its performance. In the Requirements component, we detail the requirements that LLM needs to follow when completing the specified task, including supplementary item-related information and more diverse preferences. The Standard Template and Historical Interaction Sequence components provide the standard style of LLM's response and the users' historical interaction sequence, respectively.

\begin{figure}[ht]
    \centering
    \includegraphics[width=0.7\textwidth]{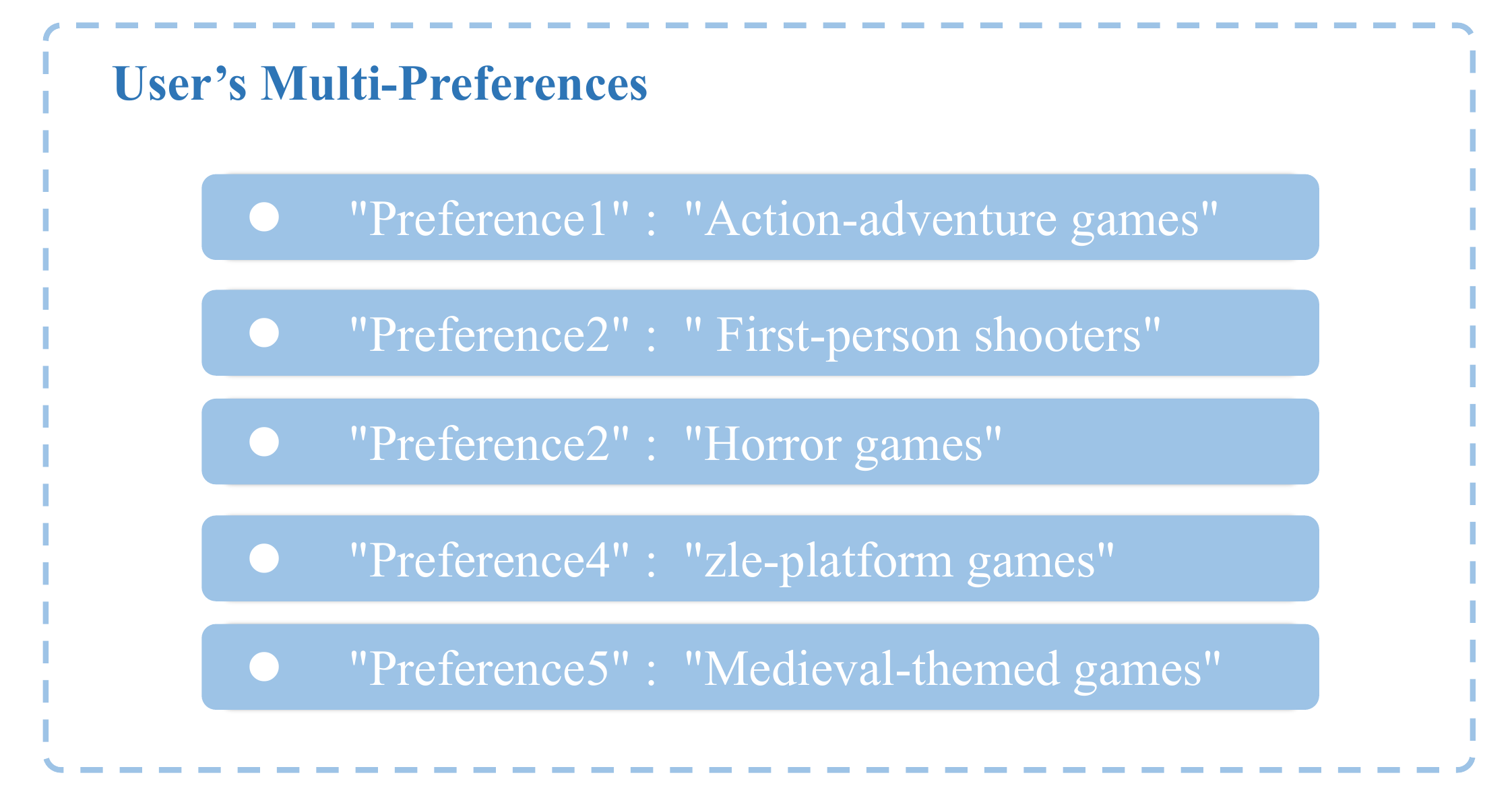}
    \caption{An example of users' multi-preferences. It was generated by GPT under the guidance of a zero-shot prompt template, where the number of user preferences \( n \) is equal to 5.}
    \label{fig:user_multi_preference}
\end{figure}

Utilizing this prompt template, we can guide the LLM model to generate users' multi-preferences, as illustrated in Figure \ref{fig:user_multi_preference}. Subsequently, by feeding the users' multiple preferences into the TextEncoder mentioned in Section \ref{3.2.2}, we can obtain embedding vectors of the users' multi-preferences. This process can be expressed by the following formulas:

\begin{equation}
\text{Preference}^u = \text{LLM}( {\text{prompt}}_p ( S^u ) ) ,
\end{equation}

\begin{equation}
\mathbf{P}^u = \text{TextEncoder}( \text{Preference}^u ) ,
\end{equation}

where \( \text{LLM}( \cdot ) \) denotes the generation of LLM model, \( {\text{prompt}}_p ( \cdot ) \) denotes the zero-shot prompt template, \( \text{Preference}^u \) denotes the \( m \) preferences of user \( u \), \( \mathbf{P}^u \in \mathbb{R}^{m \times d} \) denotes the embedding vectors of the \( m \) preferences of user \( u \).

\subsubsection{Semantic Alignment Module}
The lack of explainability in the recommendation results of the ``black box'' recommendation model stems from our inability to explain the sequence feature vectors it outputs. To achieve explainability in recommendation results, it is formally equivalent to finding an explainable alignment vector to replace the sequence feature vectors output by the ``black box'' recommender. This alignment vector would then undergo similarity calculations with the embedding vectors of candidate items to generate ranking scores. The generation of this alignment vector can be facilitated through a semantic alignment module.

\paragraph{Multi-Head Attention.}
The multi-head attention mechanism can calculates semantic similarities between queries and keys and generates corresponding attention weights based on these similarities, which are then used to weight and sum the values \cite{vaswani2017attention}. Leveraging the characteristics of multi-head attention, we treat the two sets of vectors that need alignment as queries and key-value pairs, enabling semantic alignment between these two sets of vectors.

In the multi-head attention layer, we employ the scaled dot-product attention, defined as:
\begin{equation}
\text{Attention}(\mathbf{Q}, \mathbf{K}, \mathbf{V}) = \text{softmax}\left(\frac{\mathbf{Q}\mathbf{K}^T}{\sqrt{d_k}}\right)\mathbf{V}  ,
\end{equation}
where \( \mathbf{Q} \) denotes queries, and \( \mathbf{K} \) and \( \mathbf{V} \) denotes keys and values, respectively. \( \sqrt{d_k} \) denotes the scaling factor. The multi-head attention mechanism performs the scaled dot-product attention function multiple times on \( \mathbf{Q} \), \( \mathbf{K} \), and \( \mathbf{K} \) across the \( d_k \) dimension, concatenates the outputs of these parallel single attention layers, and then performs linear projection. This can be expressed as:

\begin{equation}
\text{Multihead}(\mathbf{Q}, \mathbf{K}, \mathbf{V}) = \text{Concat}(\text{head}_1, \ldots, \text{head}_h) W^o,
\end{equation}

\begin{center}
where \( \text{head}_i = \text{Attention}(\mathbf{Q}\mathbf{W}_i^Q, \mathbf{K}\mathbf{W}_i^K, \mathbf{V}\mathbf{W}_i^V) \),
\end{center}

where \( \mathbf{W}_i^Q \in \mathbb{R}^{d \times d_k} \), \( \mathbf{W}_i^K \in \mathbb{R}^{d \times d_k} \), \( \mathbf{W}_i^V \in \mathbb{R}^{d \times d_k} \), and \( \mathbf{W}^o \in \mathbb{R}^{hd_k \times d} \) denotes projection matrices, and \( i \in \{1,2,\ldots,h\} \), \( h \) denotes the number of heads. We treat \( \mathbf{Q}^u \) as queries and \( \mathbf{P}^u \) as key-value pairs, where keys and values are the same. Leveraging the multi-head attention mechanism, we can use the users' multi-preferences embedding vector \( \mathbf{P}^u \) to capture the semantic information of the sequence feature vector \( \mathbf{Q}^u \), aligning \( \mathbf{Q}^u \) and \( \mathbf{P}^u \) semantically to obtain an alignment matrix regarding \( \mathbf{Q}^u \).

\paragraph{Position-wise Feed-Forward Networks.}
While the multi-head attention mechanism can capture local dependencies and semantic information in the input sequence, it may not be sufficient to model complex nonlinear relationships. Therefore, we introduce an additional nonlinear component by incorporating a Position-wise Feed-Forward Network to enhance the model's expressive power and learning capacity. Assuming the input vector is \( \mathbf{x} \), the definition of the Position-wise Feed-Forward Network is:

\begin{equation}
\text{FNN}(\mathbf{x}) = \text{ReLU}(\mathbf{x}\mathbf{W}_1 + \mathbf{b}_1)\mathbf{W}_2 + \mathbf{b}_2 ,
\end{equation}

where \( \mathbf{W}_1 \in \mathbb{R}^{d \times d} \), \( \mathbf{W}_2 \in \mathbb{R}^{d \times d} \) and \( \mathbf{b}_1 \in \mathbb{R}^{d} \), and \( \mathbf{b}_2 \in \mathbb{R}^{d} \) denote projection matrices and bias terms. The feedforward neural network consists of two linear transformations with a ReLU activation function between them.

\paragraph{Residual Connections and Layer Normalization.}
Residual connections allow useful low-level information to be preserved at higher layers. To better preserve the performance of the integrated model and to ensure more stable and faster training, we incorporate residual connections \cite{he2016deep}. Additionally, we utilize layer normalization to further accelerate model training and improve generalization capabilities \cite{ba2016layer}. Assuming the input vector is \( \mathbf{x} \), the definition of layer normalization is:

\begin{equation}
\text{LayerNorm}( \mathbf{x} ) = \alpha \odot \left(\frac{\mathbf{x} - \mu}{\sqrt{\sigma^2 + \epsilon}}\right) + \beta ,
\end{equation}
where \( \odot \) denotes element-wise product (Hadamard product), \( \mu \) and \( \sigma \) denote the mean and variance of \( \mathbf{x} \), \( \alpha \) and \( \beta \) denote learned scale factor and bias term.

We treat the sequence feature matrix \( \mathbf{Q}^u \in \mathbb{R}^{n \times d} \) as queries, and the users' multiple-preferences \( \mathbf{P}^u \in \mathbb{R}^{m \times d} \) as key-value pairs. The preference alignment module can be represented by the following formulas:

\begin{equation}
\text{att}^u = \text{LayerNorm}\left( \text{Multihead}(\mathbf{Q}^u, \mathbf{P}^u, \mathbf{P}^u) \right) + \mathbf{Q}^u ,
\end{equation}

\begin{equation}
\mathbf{F}^u = \left[ \begin{array}{c} \mathbf{f}_1^u \\ \mathbf{f}_2^u \\ \vdots \\ \mathbf{f}_n^u \end{array} \right] = \text{LayerNorm}\left( \text{FNN}(\text{att}^u) \right) + \text{att}^u,
\end{equation}

where \( \text{att}^u \in \mathbb{R}^{n \times d} \) denotes the result obtained after the output of the multi-head attention layer goes through layer normalization and residual connection. \( \mathbf{F}^u \in \mathbb{R}^{n \times d} \) denotes the final alignment matrix, and \( \mathbf{f}_t^u \in \mathbb{R}^d \) denotes the alignment vector corresponding to the feature vector \( \mathbf{q}_t^u \) of the subsequence composed of the first \( t \) items in user \( u \)'s interaction sequence.

\subsubsection{Prediction Module}
To prevent overfitting and reduce the number of parameters, candidate item embeddings are still obtained by retrieving the embedding matrix \( \mathbf{M} \) of embedding layer. Given the embedding vector \( \mathbf{m}_i \) of candidate item \( i \) and the alignment vector \( \mathbf{f}_t^u \), they are input into the prediction layer to obtain the recommendation score based on the feature vector \( \mathbf{q}_t^u \). The formula is as follows:

\begin{equation}
\mathbf{r}_{t,i}^u = \mathbf{f}_t^u \cdot \mathbf{m}_i ,
\end{equation}
 
where \( \mathbf{r}_{t,i}^u \) denotes the prediction score of candidate item \( i \) as the next interaction item \( \mathbf{s}_{t+1}^u \) in the user \( u \)'s interaction sequence \( \{ \mathbf{s}_1^u, \mathbf{s}_2^u, \ldots, \mathbf{s}_t^u\} \).

\paragraph{Model Training.}
As mentioned in the Section \ref{3.2.2}, We transform the interaction sequence \( \{S_1^u, S_2^u, \ldots, S_{|S^u| - 1}^u\} \) of user \( u \) into a fixed-length vector sequence \( \{ \mathbf{s}_1^u, \mathbf{s}_2^u, \ldots, \mathbf{s}_n^u\} \), where \( n \) is a fixed length. We denote the collection of all user vector sequences as \( \mathbf{s} \), where \( \mathbf{s}^u \in \mathbf{s} \). Suppose given \( \{\mathbf{s}_1^u, \mathbf{s}_2^u, \ldots, \mathbf{s}_t^u\} \), we denote its next expected item \( \mathbf{s}_{t+1}^u \) as the positive sample \( \mathbf{pos}_t \). A negative sample is randomly sampled from the item set \( I \), and its embedding vector is denoted as \( \mathbf{neg}_t \), where \( \mathbf{neg}_t \notin \mathbf{s}^u \). Binary cross-entropy loss is used as the loss function:
\begin{equation}
\mathcal{L} = - \sum_{\mathbf{s}^u \in \mathbf{s}} \sum_{t \in \{1, 2, ..., n\}} [ \log(\sigma(\mathbf{r}_{t,\mathbf{pos}_t}^u)) + \log(1 - \sigma(\mathbf{r}_{t,\mathbf{neg}_t}^u))] .
\end{equation}

The model is optimized by the Adam optimizer, which is a variant of stochastic gradient descent (SGD) with adaptive moment estimation \cite{kingma2014adam}. To prevent overfitting, we also adopt the Early Stopping strategy. When the performance of the model on the validation set stabilizes and no longer improves, we terminate the training in advance.

\begin{figure}[t]
    \centering
    \includegraphics[width=0.9\textwidth]{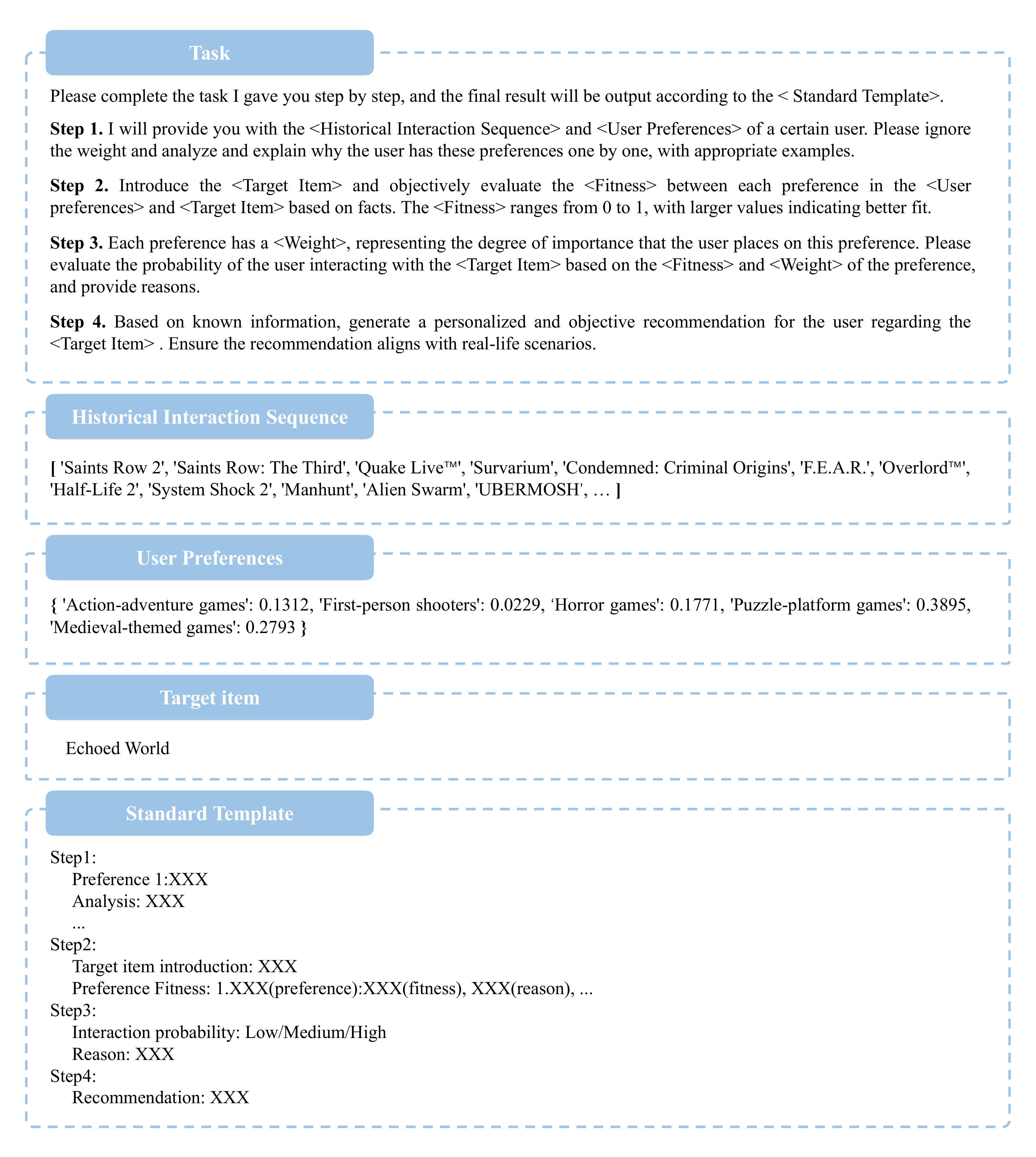}
    \caption{The CoT prompt template. It mainly consists of four progressive steps, and needs to fill in the user's interaction sequence, target item, user‘s multi-preferences and attention weight, also taking the data on the Steam dataset as an example.}
    \label{fig:CoT_prompt}
\end{figure}
\subsubsection{Explainable Recommendation Generation Module}
In this module, we guide the LLM model to generate explainable reasons for our recommendation results. Since the generated recommendation texts aim to provide semantic explanations for the recommendation model's results from the perspective of user preferences and there are no expected recommendation texts, it is not involved in model training but serves as an output head to provide explanations for the recommendation results.

As described in the previous sections, by inputting \( \{ \mathbf{s}_1^u, \mathbf{s}_2^u, \ldots, \mathbf{s}_n^u\} \) into the model, we can obtain the feature vector \( \mathbf{q}_n^u \) of this sequence, as well as the users' multi-preferences \( \text{preference}^u \) and its embedding \( \mathbf{P}^u \). By using the projection matrices in the preference alignment module, we can obtain the attention weights regarding users' multi-preferences, which can be described as follows:

\begin{equation}
\mathbf{Q}_n^u = \text{Concat}(\mathbf{q}_n^u \mathbf{W}_1^Q, \ldots, \mathbf{q}_n^u \mathbf{W}_h^Q),
\end{equation}

\begin{equation}
\mathbf{K}^u = \text{Concat}(\mathbf{P}^u \mathbf{W}_1^K, \ldots, \mathbf{P}^u \mathbf{W}_h^K),
\end{equation}

\begin{equation}
\mathbf{\omega}^u = \text{softmax}\left(\frac{{\mathbf{Q}_n^u \mathbf{K}^u}^T}{\sqrt{{hd}_k}}\right),
\end{equation}

where \( \mathbf{W}_i^Q \in \mathbb{R}^{d \times d_k} \), \( \mathbf{W}_i^K \in \mathbb{R}^{d \times d_k} \) are the projection matrices in the preference alignment module, \( \mathbf{Q}_n^u \in \mathbb{R}^{1 \times hd_k} \), \( \mathbf{K}^u \in \mathbb{R}^{m \times hd_k} \) denote the projected \( \mathbf{q}_n^u \) and \( \mathbf{P}^u \) respectively, and \( \mathbf{\omega}^u \in \mathbb{R}^{1 \times m} \) denotes the attention weights of the users' multi-preferences. The calculated attention weights would be optimized to select the most aligned LLM's generated preference and recommender's (e.g. SASRec) preference embedding within the training process of attention module.

The Chain-of-Thought (CoT) prompt leveraging intermediate reasoning steps to enable LLM to achieve complex reasoning capabilities \cite{wei2022chain}. To guide LLM accurately in generating the explainable recommendation texts we desire, we have meticulously designed a zero-shot CoT prompt template, which consists of four progressive steps, as illustrated in Figure \ref{fig:CoT_prompt}.

By employing these four progressive steps, we guide LLM to achieve the ultimate goal of generating explainable recommendation text. This module can be described as:

\begin{equation}
\text{Explanation}^u = \text{LLM}(\text{prompt}_{e}(S^u, \text{preference}^u, \mathbf{\omega}^u, S_{|S^u|}^u))  ,
\end{equation}

where \( \text{prompt}_{e}(\cdot) \) denotes the zero-shot CoT prompt template. \( S^u = \{S_1^u, S_1^u, ..., S_{|S^u|-1}^u\} \) denotes the interaction sequence of user u. \( S_{|S^u|}^u \) denotes the target item. \( \text{Explanation}^u \) denotes the explainable recommendation texts.

\section{Experiments}
\subsection{Experiment Settings}
\label{4.1}
\paragraph{Dateset.}
We evaluated our method on three real-world datasets, which exhibit significant differences in domain and sparsity. These datasets have unique and distinct item titles, enabling LLMs to utilize only item titles to understand the relevant information and generate explainable recommendations.

\begin{itemize}
    \item \textbf{MovieLens:} MovieLens is a classic dataset for movie recommendation systems, created and maintained by the GroupLens Research lab \cite{harper2015movielens}. We used the version with 1 million user ratings (ML-1M).
    \item \textbf{Amazon:} Amazon is one of the world's largest online retailers, selling a variety of products including cosmetics and books. We used the large-scale Amazon review dataset collected by the McAuley Lab — Amazon Reviews 2014 \cite{mcauley2015image}. This dataset is divided into several individual datasets according to top product categories on Amazon. We adopted the Beauty category.
    \item \textbf{Steam:} Steam is one of the world's largest digital game distribution platforms, offering game purchases, social networking, digital rights management, and more. We used the Steam dataset introduced by Kang et al. \cite{kang2018self}, which includes user reviews and game information scraped from the Steam platform.
\end{itemize}

We regard the presence of reviewer ratings as implicit feedback (i.e., user-item interactions) and use timestamps to determine the sequence of actions. To avoid excessive data sparsity and improve recommendation quality, we discarded users and items with fewer than 5 interactions in ML-1M and Beauty datasets, and fewer than 20 interactions in the Steam dataset. Additionally, we divided each user's historical sequence \( S^u \) into three parts based on their usage: (1). the most recent action \( S_{|S^u|}^u \)  for testing, (2). the second most recent action \( S_{|S^u|-1}^u \)  for validation, and (3). all remaining actions for training. Finally, we also removed a small number of user sequences from which large models could not correctly extract preferences.

\begin{table}[t]
  \caption{Statistics of the datasets}
  \label{tab:statistics}
  \centering
  \begin{tabular}{cccc}
    \toprule
    \textbf{Dataset}   & \textbf{Beauty}    & \textbf{Steam}     & \textbf{ML-1M}   \\
    \midrule
    \#Users             & 21849	    & 39626	    & 6040    \\
    \#Items             & 12066	    & 9261	    & 3416    \\
    \#actions           & 195141    & 1774231   & 999611  \\
    Avg.actions/User    & 8.93	    & 44.77	    & 163.5   \\
    Avg.actions/Item    & 16.17	    & 191.58	& 292.63  \\
    Sparsity            & 99.93\%   & 99.52\%   & 95.16\% \\
    \bottomrule
  \end{tabular}
\end{table}

After preprocessing, the statistics of the datasets is shown in Table \ref{tab:statistics}. The Beauty dataset has the smallest average number of interactions per user and per item, making it the sparsest. The Steam dataset follows, while the ML-1M dataset is the most dense.

\paragraph{Baseline.}
We selected three widely used sequential recommendation models as baseline models:

\begin{itemize}
    \item \textbf{GRU4Rec} \cite{hidasi2015session} based on Gated Recurrent Unit (GRU) architecture, learns representations of user sequence behaviors for personalized recommendations. It possesses the ability to capture long-term dependencies and handle variable-length sequences.
    \item \textbf{BERT4Rec} \cite{sun2019bert4rec} leverages pre-trained BERT models to achieve deeper semantic understanding and personalized recommendations by learning representations of user historical behavior sequences.
    \item \textbf{SASRec} \cite{kang2018self} is a sequential recommendation model based on self-attention mechanism. By introducing self-attention mechanism, it effectively captures the correlation between different items in user behavior sequences.
\end{itemize}

By integrating these three baseline models into our framework, we obtain LANE-GRU4Rec, LANE-BERT4Rec, and LANE-SASRec, respectively.

\paragraph{Evaluation Metrics.} To accurately evaluate the performance differences between recommendation models, we adopted two common Top-N metrics: HitRate@10 and NDCG@10. To avoid the computational burden of evaluating all user-item pairs, we followed the evaluation strategies described in \cite{kang2018self} and \cite{he2017neural}. For each user \( u \), we randomly sampled 100 negative items and ranked these items along with the ground truth item. Based on the ranking of these 101 items, we evaluated the performance using HitRate@10 and NDCG@10.

\paragraph{Implementation Details.} \label{details}To ensure a fair comparison, all baseline models and the proposed framework were implemented using the PyTorch framework and optimized using the Adam optimizer. Other hyperparameters and initialization strategies were kept the same as in the original papers or provided by the open-source code of the respective models.
For the proposed framework, we integrated baseline models as the recommendation model to be explained. Our framework utilizes GPT-3.5 as the large language model for generating multiple preferences and transcribing recommendation reasons. The embedding dimension \( d \) was set to 384 (the same as Sentence-BERT), the number of heads \( h \) was 4, the hidden size \( d_k \) was 384, the number of user preferences \( m \) was 5, the learning rate was 0.001, the batch size was 128, and the dropout rate was 0.5. For the ML-1M dataset, we set the maximum sequence length \( n \) to 200, and for the other two datasets, the maximum sequence length \( n \) was set to 50. All other hyperparameters used within baseline models were consistent with those in the original paper.

\subsection{Experiment Results}
Our proposed explainable recommendation framework is highly flexible and can be integrated with any type of sequential recommendation model. Table \ref{tab:recommendation performance comparison} shows the recommendation performance of our proposed explainable framework and all baseline models on three datasets. Through the performance results of the baseline model before and after nested frameworks, we can see that our proposed frameworks have achieved significant performance improvements on all baseline models, which further confirms the effectiveness of our model in enhancing traditional recommendations. Compared with the original baseline model, LANE-GRU4Rec's NDCG@10 and HitRate@10 increased by 7.52\% and 4.51\% on average, LANE-BERT4Rec's NDCG@10 and HitRate@10 increased by 12.44\% and 9.67\% on average, and LANE-SASRec's NDCG@10 and HitRate@10 increased by 15.09\% and 11.37\% on average. This performance improvement is attributed to the stronger semantic derivation and induction capabilities of LLMs than the original baseline model. Based on their rich knowledge, LLMs can deduce additional higher-level semantic information from interaction sequences, thereby achieving stronger recommendation performance than integrated models.

\begin{table}[ht]
\label{tab:recommendation performance comparison}
\centering
\caption{Recommendation Performance Comparison}
\resizebox{.98\textwidth}{!}{
\begin{tabular}{l|cccccc}
\toprule
\multirow{2}{*}{\centering \textbf{Dataset}} & \multicolumn{2}{c}{\textbf{Beauty}} & \multicolumn{2}{c}{\textbf{Steam}} & \multicolumn{2}{c}{\textbf{ML-1M}} \\
\cmidrule(lr){2-3} \cmidrule(lr){4-5} \cmidrule(lr){6-7}
 & \textbf{NDCG@10} & \textbf{HR@10} & \textbf{NDCG@10} & \textbf{HR@10} & \textbf{NDCG@10} & \textbf{HR@10} \\
\midrule
GRU4Rec & 0.2853 & 0.4422 & 0.5025 & 0.7492 & 0.5277 & 0.7614 \\
LANE-GRU4Rec & 0.3238 & 0.4825 & 0.5109 & 0.7598 & 0.5667 & 0.7844 \\
\cmidrule(lr){1-7}
\textbf{Improv.} & \textbf{13.49\%} & \textbf{9.11\%} & \textbf{1.67 \%} & \textbf{1.41\%} & \textbf{7.39\%} & \textbf{3.02\%} \\
\cmidrule(lr){1-7}
BERT4Rec & 0.224 & 0.3808 & 0.477 & 0.7261 & 0.4611 & 0.7151 \\
LANE-ERT4Rec & 0.2734 & 0.4526 & 0.4982 & 0.7495 & 0.5111 & 0.7646 \\
\cmidrule(lr){1-7}
\textbf{Improv.} & \textbf{22.05\%} & \textbf{18.86\%} & \textbf{4.44\%} & \textbf{3.22\%} & \textbf{10.84\%} & \textbf{6.92\%} \\
\cmidrule(lr){1-7}
SASRec & 0.2831 & 0.423 & 0.4789 & 0.728 & 0.5701 & 0.7983 \\
LANE-SASRec & 0.3511 & 0.5172 & 0.5649 & 0.803 & 0.5888 & 0.8106 \\
\cmidrule(lr){1-7}
\textbf{Improv.} & \textbf{24.02\%} & \textbf{22.27\%} & \textbf{17.96\%} & \textbf{10.30\%} & \textbf{3.28\%} & \textbf{1.54\%} \\
\bottomrule
\end{tabular}}
\end{table}

In addition, we can find that the improvement achieved on sparse datasets is more obvious than that on dense datasets. For example, on the Beauty dataset, LANE-SASRec improved NDCG@10 and HR@10 by 24.02\% and 22.27\% over SASRec, but on ML-1M dataset, the improvement was only 3.28\% and 1.54\%.Because in sparse data sets, the features that can be extracted by the original sequence recommendation model are limited, the improvement brought by the additional semantic information brought by LLMs will be more obvious.

\subsection{Sensitivity Analysis}
We conducted sensitivity analysis to investigate the impact of four important hyperparameters: hidden size\( d_k \), number of heads \( h \), maximum sequence length \( n \), and number of user preferences \( m \). In each experiment, we only changed the current hyperparameter under study while keeping the remaining hyperparameters at their default values as specified in Section \ref{4.1}. Additionally, we selected HitRate@5, HitRate@10, NDCG@5, and NDCG@10 as evaluation metrics and performed all hyperparameter experiments on the ML-1M dataset. 

\begin{figure}[ht]
    \centering
    \includegraphics[width=0.98\textwidth]{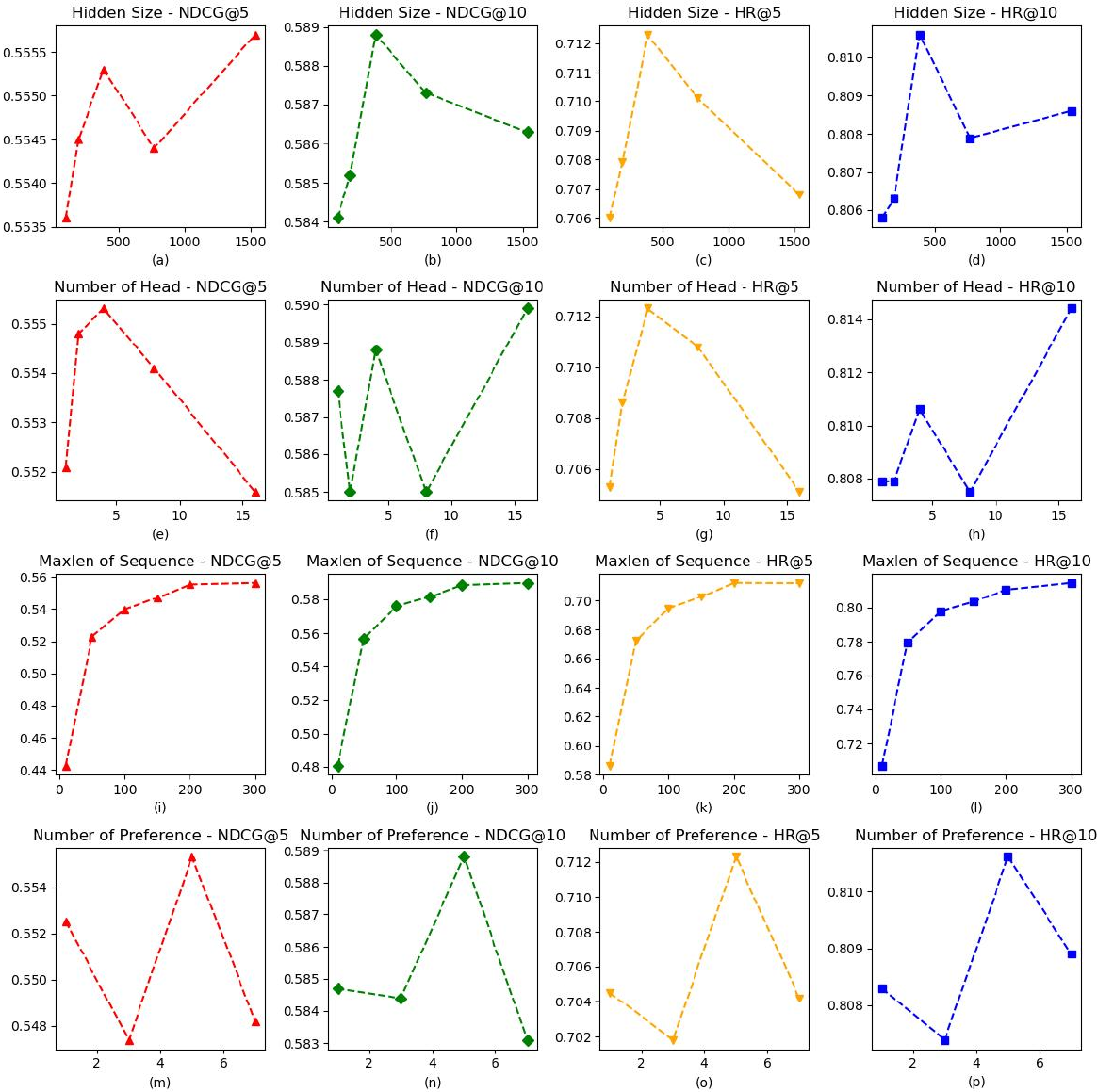}
    \caption{The experimental results of the sensitivity analysis on the ML-1M dataset for the four hyperparameters: (a) - (d) hidden size\( d_k \), (e) - (h) number of heads \( h \), (i) - (l) maximum sequence length \( n \), (m) - (p) and number of user preferences \( m \). The evaluation metrics used are NDCG@5, NDCG@10, HitRate@5, and HitRate@10, .}
    \label{fig:hyperparameters}
\end{figure}

The specific experimental results are shown in Figure \ref{fig:hyperparameters}. According to Figure \ref{fig:hyperparameters} (a) - (d), it is observed that when the hidden size \( d_k \) equals the embedding dimension \( d \), the model performs optimally. Moreover, within a certain range, increasing \( d_k \) leads to better model performance, but excessively large values may result in overfitting.
Figure \ref{fig:hyperparameters} (e) - (h) indicate that appropriately increasing the number of heads \( h \) appropriately can enhance the model's projection capability, leading to better generalization but sacrificing some accuracy. Consequently, in Top-N recommendations, when N is relatively large, the recommendation performance improves. However, when N is small, the performance may deteriorate.
Figure \ref{fig:hyperparameters} (i) - (l) indicate that increasing the maximum sequence length \( n \) improves recommendation performance, but the effect gradually diminishes. Considering the balance between performance and training cost, it is common to choose a value slightly larger than the average sequence length of users. 
As for the number of user preferences \( m \), Figure \ref{fig:hyperparameters} (m) - (p) indicate that the performance is optimal when \(m=5\). \( m \) should be chosen moderately, as too large a value may lead to information loss, while too small a value may introduce noise and increase training costs.

\subsection{Case Study}
In this study, we randomly selected a user interaction sequence from the Steam dataset as a sample to analyze the explanation results output by our framework (Figure \ref{fig:case_study}). The complete results are shown in Appendix \ref{appendix:case study}. According to the CoT prompt and the standard response template we specified, LLM generates the corresponding text step by step. The core idea is to simulate and replicate the recommendation process of the recommendation model for explanation purposes.

\begin{figure}[ht]
    \centering
    \includegraphics[width=1\textwidth]{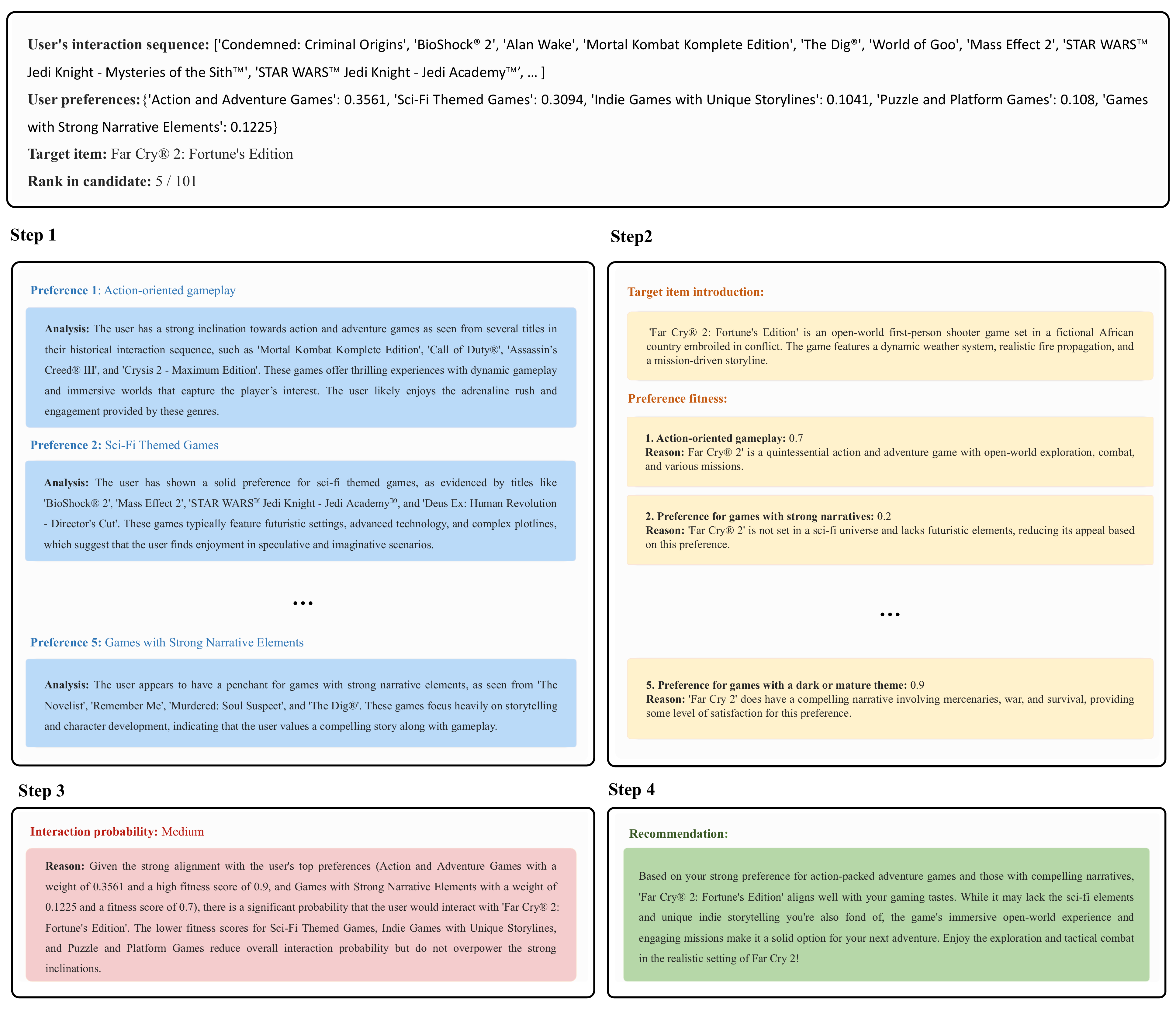}
    \caption{Recommended results and corresponding explanations for the sample. \textbf{Top}: "User preferences" gives the user preference and the corresponding attention weight, and "Rank in candidate" is the ranking of the target items finally output by our model. \textbf{Bottom}: The explanation generated by LLM under the guidance of the CoT prompt template. It gives the results of the four steps in the CoT prompt template respectively.}
    \label{fig:case_study}
\end{figure}

% In Step 1, the model simulates the process of extracting features from the user interaction sequence. Step 2 and Step 3 simulate the embedding of the target item into vectors and the calculation of the recommendation score, respectively. As for Step 4, its function is to transform the process-based explanation obtained in the previous steps into personalized recommendation statements that fit real-world scenarios.
Step 1 simulates and makes transparent the process of extracting features from user interaction sequences. In this step, our framework guides LLM to explain the origin of the previously extracted user interaction sequence features (user's multiple preferences). For example, for the origin of the user's preference for "Action-oriented gameplay", LLM explains that the user has also played games with intense and fast-paced action elements such as `Mortal Kombat Komplete Edition', `Call of Duty®', `Assassin's Creed® III' and `Crysis 2 - Maximum Edition', and speculates that the user may like the adrenaline rush and sense of participation provided by these types of games.

Step 2 simulates and makes transparent the process of embedding the target item. In this step, LLM will first generate a basic introduction for the target item, which provides users with basic information about the recommended item while ensuring that LLM can grasp the relevant information of the target item. Subsequently, LLM will compare the target item information it has grasped with the user's multiple preferences one by one, evaluate the fit between the preferences and the target item, and explain the reasons. For example, for the good article "Action-oriented gameplay", the fit evaluation generated by LLM is "0.7", and the explanation given is "Far Cry® 2 is a typical action-adventure game with open world exploration, combat, and various missions."

Step 3 simulates and makes transparent the process of obtaining the recommendation score. In this step, LLM will predict whether the user will interact with the target item based on the fit evaluation of each preference in Step 2 and the attention weight of the user's multiple preferences, and give a corresponding explanation. In the example we gave, LLM gave an interaction probability evaluation of ``Medium'' and gave the reason for the evaluation in "Reason".

The role of Step 4 is to convert the process-based explanations obtained in the previous steps into personalized recommendation text that fits the real scene. In this step, GPT will combine all the information in the first three steps to generate a personalized recommendation text about the target item for the user. From the generated recommendation text, we can see that the target item fits the user's preferences, such as "realistic combat", "intense action" and "engaging and immersive experience".

\subsection{Quality Analysis of Explanation}
We used an expert scoring method to evaluate the quality of the explanations generated by our model from seven metrics: clarity, detail, effectiveness, relevance, logic, trust, and satisfaction. The specific meaning of each metric is described by a question and scored on a 5-point scale. The model is anonymous. The complete questions are shown in Appendix \ref{appendix:seven metrics}. We selected ERNIE-4.0, GPT-3.5-Turbo, and GPT-4o as the baseline models for this experiment. 

In order to distinguish the different needs of merchants and users for the explanation content output by the recommendation system, we conducted two surveys: 1). Survey 1 is for user(consumer)-oriented explanations. Compared with the lengthy explanation of the model recommendation process, users pay more attention to the explanation of the recommendation results. Therefore, we randomly selected 50 samples from the Steam dataset and only intercepted the recommendation output by the model in Step 4 as the evaluation object. For fair comparison, we also let the other baseline models generate explanations of comparable text length and avoid outputting redundant process analysis. 2). Survey 2 is for merchant-oriented explanations. Unlike users, merchants need more detailed recommendation explanations, including explanations of the recommendation process of the recommendation system. Therefore, we randomly selected 20 samples and retained the complete explanation of the model output as the evaluation object. Correspondingly, we also let the baseline model give its complete recommendation process.

\begin{figure}[ht]
    \centering
    \includegraphics[width=1\textwidth]{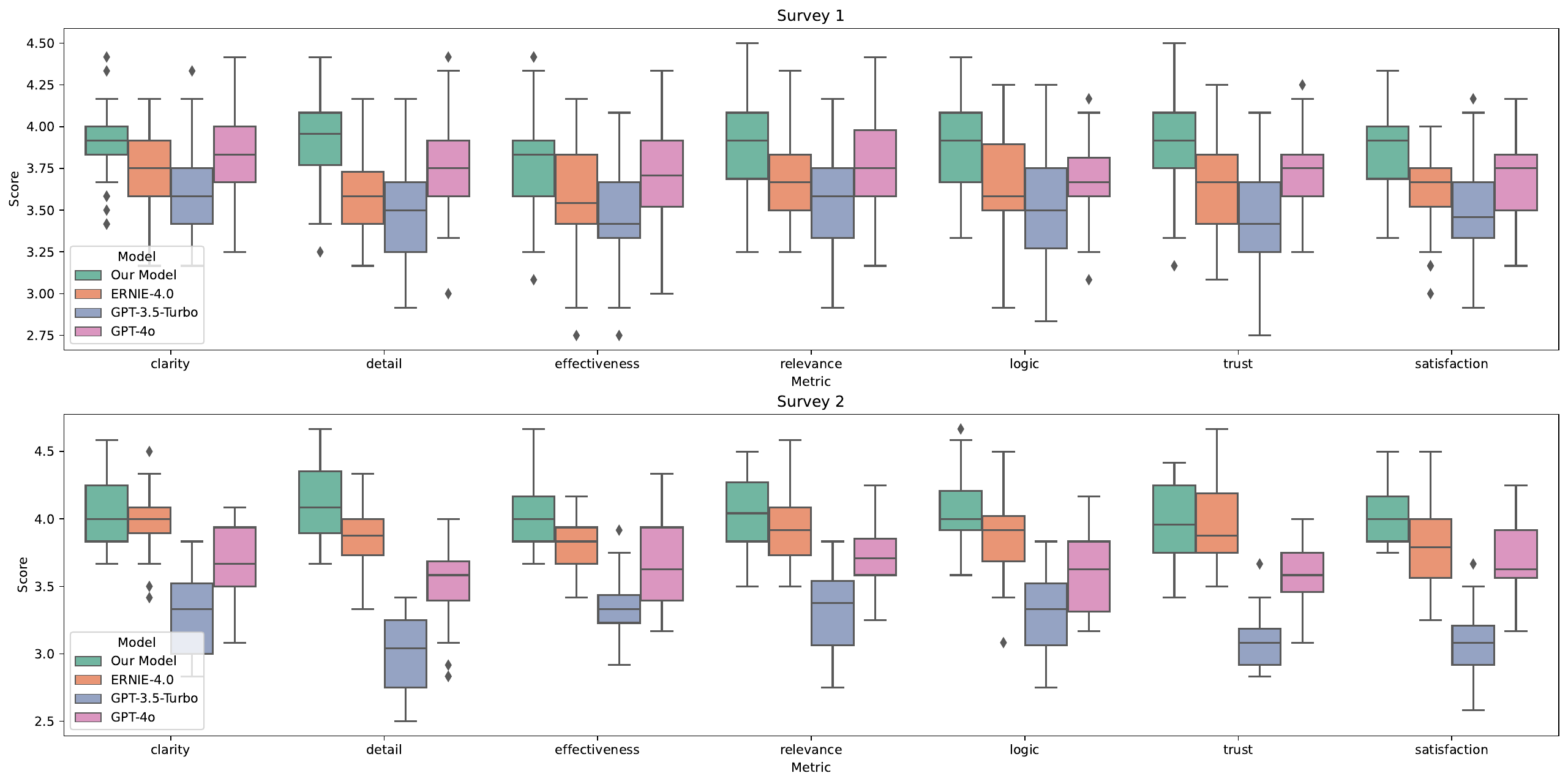}
    \caption{The score distribution of all samples for each model in Survey 1 and Survey 2 on the seven metrics.}
    \label{fig:scores_distribution}
\end{figure}
\begin{figure}[t]
    \centering
    \includegraphics[width=1\textwidth]{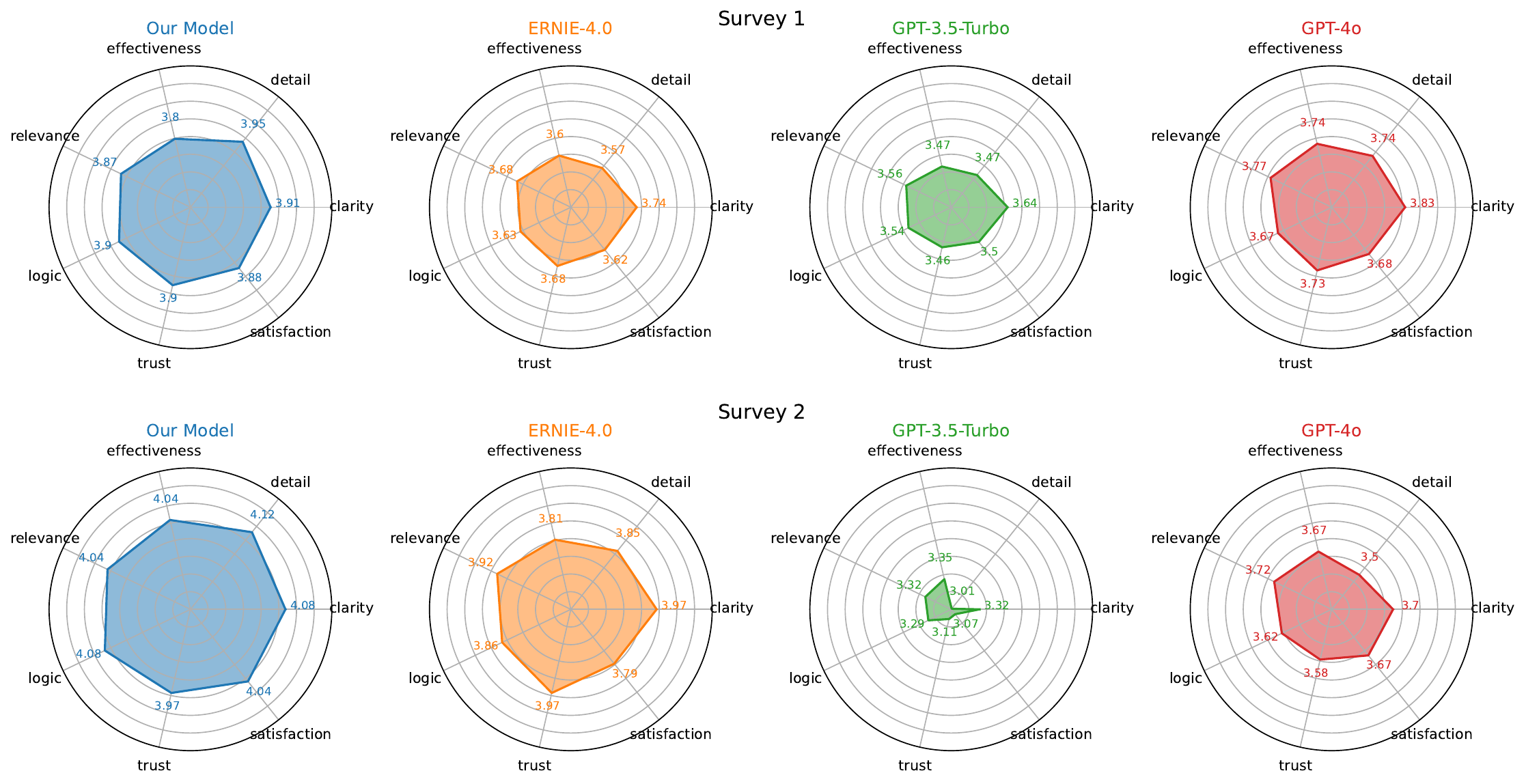}
    \caption{The average scores of all samples for each model in Survey 1 and Survey 2 on the seven metrics.}
    \label{fig:average_scores_radar_chart}
\end{figure}

The score distribution of all samples in Survey 1 and Survey 2 is shown in Figure \ref{fig:scores_distribution}. We can see that our model is generally above other models in the distribution of the seven metrics, and performs best. Specifically, our model is significantly better than other models in terms of detail, relevance, logic, and satisfaction, and the score distribution of each metric is relatively concentrated, showing stable high performance. "ERNIE-4.0" and "GPT-4o" performed second, each with its own advantages, and only in some metrics the score distribution was wide and showed fluctuations. "GPT-3.5-turbo" scored the lowest in all metrics, and the distribution was relatively scattered, indicating that its performance was poor and not stable enough.

The average scores of all samples in survey 1 and survey 2 for each metric are shown in Figure \ref{fig:average_scores_radar_chart}. We can see that our model has the best average scores on the 7 metrics in both surveys, "GPT-4.0" and "ERNIE-4.0" are in the middle, and GPT-3.5-Turbo is the weakest. Among them, "GPT-4.0" performs better than "ERNIE-4.0" in survey 1, and vice versa in survey 2.

\section{Conclusions}
% we proposes an explainable recommendation framework based on LLMs. It is a model-agnostic explainable recommendation method.
% Without the need for fine-tuning LLMS, it can enhance recommendation performance based on its integrated traditional `black-box' recommendation models, while generating personalized explanations according to user preferences.
% The framework utilizes items' titles to achieve semantic embedding of items, extracts user preferences from sequences using zero-shot prompts, and uses a multi-head attention mechanism to semantically align the feature vectors output by the integrated model with user preferences. Finally, it provides explanations by simulating the recommendation process using CoT prompts.
% We conducted experiments on several real-world benchmark datasets to validate the effectiveness of the framework, and through a questionnaire survey, we demonstrated its ability to generate accurate and diverse high-quality explanations, with high user satisfaction.
We propose an innovative explainable recommendation framework based on large language models (LLMs). This framework can improve the recommendation performance of its integrated "black box" recommendation model without tuning parameter-complex LLMs, and utilizes the language generation ability of LLMs to generate comprehensive and highly interpretable recommendation logic. Therefore, many more powerful closed-source commercial large language models can also enhance the explainability of online recommendation systems through API calls. Our research highlights the potential of LLMs as explainers in recommendation systems and their advantages in reducing training and maintenance costs. We conducted experiments on several real-world benchmark datasets to verify the effectiveness of the framework, and demonstrated its ability to generate accurate, diverse, high-quality explanations and obtain high user satisfaction through visualization cases and questionnaire voting. Future work can further explore how to further optimize this framework to adapt to different types and sizes of recommendation systems, provide more diversified explanations, and expand to a wider range of commercial applications.

\bibliographystyle{acm}
\bibliography{LANE}

\begin{thebibliography}{10}

\bibitem{achiam2023gpt}
{\sc Achiam, J., Adler, S., Agarwal, S., Ahmad, L., Akkaya, I., Aleman, F.~L., Almeida, D., Altenschmidt, J., Altman, S., Anadkat, S., et~al.}
\newblock Gpt-4 technical report.
\newblock {\em arXiv preprint arXiv:2303.08774\/} (2023).

\bibitem{ba2016layer}
{\sc Ba, J.~L., Kiros, J.~R., and Hinton, G.~E.}
\newblock Layer normalization.
\newblock {\em arXiv preprint arXiv:1607.06450\/} (2016).

\bibitem{balog2019transparent}
{\sc Balog, K., Radlinski, F., and Arakelyan, S.}
\newblock Transparent, scrutable and explainable user models for personalized recommendation.
\newblock In {\em Proceedings of the 42nd international acm sigir conference on research and development in information retrieval\/} (2019), pp.~265--274.

\bibitem{bao2023tallrec}
{\sc Bao, K., Zhang, J., Zhang, Y., Wang, W., Feng, F., and He, X.}
\newblock Tallrec: An effective and efficient tuning framework to align large language model with recommendation.
\newblock In {\em Proceedings of the 17th ACM Conference on Recommender Systems\/} (2023), pp.~1007--1014.

\bibitem{brown2020language}
{\sc Brown, T., Mann, B., Ryder, N., Subbiah, M., Kaplan, J.~D., Dhariwal, P., Neelakantan, A., Shyam, P., Sastry, G., Askell, A., et~al.}
\newblock Language models are few-shot learners.
\newblock {\em Advances in neural information processing systems 33\/} (2020), 1877--1901.

\bibitem{chen2018attention}
{\sc Chen, J., Zhuang, F., Hong, X., Ao, X., Xie, X., and He, Q.}
\newblock Attention-driven factor model for explainable personalized recommendation.
\newblock In {\em The 41st international ACM SIGIR conference on research \& development in information retrieval\/} (2018), pp.~909--912.

\bibitem{chen2019personalized}
{\sc Chen, X., Chen, H., Xu, H., Zhang, Y., Cao, Y., Qin, Z., and Zha, H.}
\newblock Personalized fashion recommendation with visual explanations based on multimodal attention network: Towards visually explainable recommendation.
\newblock In {\em Proceedings of the 42nd International ACM SIGIR Conference on Research and Development in Information Retrieval\/} (2019), pp.~765--774.

\bibitem{cheng2023explainable}
{\sc Cheng, H., Wang, S., Lu, W., Zhang, W., Zhou, M., Lu, K., and Liao, H.}
\newblock Explainable recommendation with personalized review retrieval and aspect learning.
\newblock {\em arXiv preprint arXiv:2306.12657\/} (2023).

\bibitem{cheng2019incorporating}
{\sc Cheng, W., Shen, Y., Huang, L., and Zhu, Y.}
\newblock Incorporating interpretability into latent factor models via fast influence analysis.
\newblock In {\em Proceedings of the 25th ACM SIGKDD International Conference on Knowledge Discovery \& Data Mining\/} (2019), pp.~885--893.

\bibitem{devlin2018bert}
{\sc Devlin, J., Chang, M.-W., Lee, K., and Toutanova, K.}
\newblock Bert: Pre-training of deep bidirectional transformers for language understanding.
\newblock {\em arXiv preprint arXiv:1810.04805\/} (2018).

\bibitem{dong2022survey}
{\sc Dong, Q., Li, L., Dai, D., Zheng, C., Wu, Z., Chang, B., Sun, X., Xu, J., and Sui, Z.}
\newblock A survey on in-context learning.
\newblock {\em arXiv preprint arXiv:2301.00234\/} (2022).

\bibitem{fan2023recommender}
{\sc Fan, W., Zhao, Z., Li, J., Liu, Y., Mei, X., Wang, Y., Tang, J., and Li, Q.}
\newblock Recommender systems in the era of large language models (llms).
\newblock {\em arXiv preprint arXiv:2307.02046\/} (2023).

\bibitem{gedikli2014should}
{\sc Gedikli, F., Jannach, D., and Ge, M.}
\newblock How should i explain? a comparison of different explanation types for recommender systems.
\newblock {\em International Journal of Human-Computer Studies 72}, 4 (2014), 367--382.

\bibitem{guan2024enhancing}
{\sc Guan, Z., Wu, L., Zhao, H., He, M., and Fan, J.}
\newblock Enhancing collaborative semantics of language model-driven recommendations via graph-aware learning.
\newblock {\em arXiv preprint arXiv:2406.13235\/} (2024).

\bibitem{guan2024langtopo}
{\sc Guan, Z., Zhao, H., Wu, L., He, M., and Fan, J.}
\newblock Langtopo: Aligning language descriptions of graphs with tokenized topological modeling.
\newblock {\em arXiv preprint arXiv:2406.13250\/} (2024).

\bibitem{harper2015movielens}
{\sc Harper, F.~M., and Konstan, J.~A.}
\newblock The movielens datasets: History and context.
\newblock {\em Acm transactions on interactive intelligent systems (tiis) 5}, 4 (2015), 1--19.

\bibitem{he2016deep}
{\sc He, K., Zhang, X., Ren, S., and Sun, J.}
\newblock Deep residual learning for image recognition.
\newblock In {\em Proceedings of the IEEE conference on computer vision and pattern recognition\/} (2016), pp.~770--778.

\bibitem{he2015trirank}
{\sc He, X., Chen, T., Kan, M.-Y., and Chen, X.}
\newblock Trirank: Review-aware explainable recommendation by modeling aspects.
\newblock In {\em Proceedings of the 24th ACM international on conference on information and knowledge management\/} (2015), pp.~1661--1670.

\bibitem{he2017neural}
{\sc He, X., Liao, L., Zhang, H., Nie, L., Hu, X., and Chua, T.-S.}
\newblock Neural collaborative filtering.
\newblock In {\em Proceedings of the 26th international conference on world wide web\/} (2017), pp.~173--182.

\bibitem{he2023large}
{\sc He, Z., Xie, Z., Jha, R., Steck, H., Liang, D., Feng, Y., Majumder, B.~P., Kallus, N., and McAuley, J.}
\newblock Large language models as zero-shot conversational recommenders.
\newblock In {\em Proceedings of the 32nd ACM international conference on information and knowledge management\/} (2023), pp.~720--730.

\bibitem{heckel2017scalable}
{\sc Heckel, R., Vlachos, M., Parnell, T., and D{\"u}nner, C.}
\newblock Scalable and interpretable product recommendations via overlapping co-clustering.
\newblock In {\em 2017 IEEE 33rd International Conference on Data Engineering (ICDE)\/} (2017), IEEE, pp.~1033--1044.

\bibitem{hidasi2015session}
{\sc Hidasi, B., Karatzoglou, A., Baltrunas, L., and Tikk, D.}
\newblock Session-based recommendations with recurrent neural networks.
\newblock {\em arXiv preprint arXiv:1511.06939\/} (2015).

\bibitem{hou2024large}
{\sc Hou, Y., Zhang, J., Lin, Z., Lu, H., Xie, R., McAuley, J., and Zhao, W.~X.}
\newblock Large language models are zero-shot rankers for recommender systems.
\newblock In {\em European Conference on Information Retrieval\/} (2024), Springer, pp.~364--381.

\bibitem{huang2023recommender}
{\sc Huang, X., Lian, J., Lei, Y., Yao, J., Lian, D., and Xie, X.}
\newblock Recommender ai agent: Integrating large language models for interactive recommendations.
\newblock {\em arXiv preprint arXiv:2308.16505\/} (2023).

\bibitem{kang2018self}
{\sc Kang, W.-C., and McAuley, J.}
\newblock Self-attentive sequential recommendation.
\newblock In {\em 2018 IEEE international conference on data mining (ICDM)\/} (2018), IEEE, pp.~197--206.

\bibitem{kingma2014adam}
{\sc Kingma, D.~P., and Ba, J.}
\newblock Adam: A method for stochastic optimization.
\newblock {\em arXiv preprint arXiv:1412.6980\/} (2014).

\bibitem{kojima2022large}
{\sc Kojima, T., Gu, S.~S., Reid, M., Matsuo, Y., and Iwasawa, Y.}
\newblock Large language models are zero-shot reasoners.
\newblock {\em Advances in neural information processing systems 35\/} (2022), 22199--22213.

\bibitem{li2023personalized}
{\sc Li, L., Zhang, Y., and Chen, L.}
\newblock Personalized prompt learning for explainable recommendation.
\newblock {\em ACM Transactions on Information Systems 41}, 4 (2023), 1--26.

\bibitem{lin2024data}
{\sc Lin, X., Wang, W., Li, Y., Yang, S., Feng, F., Wei, Y., and Chua, T.-S.}
\newblock Data-efficient fine-tuning for llm-based recommendation.
\newblock {\em arXiv preprint arXiv:2401.17197\/} (2024).

\bibitem{liu2023chatgpt}
{\sc Liu, J., Liu, C., Zhou, P., Lv, R., Zhou, K., and Zhang, Y.}
\newblock Is chatgpt a good recommender? a preliminary study.
\newblock {\em arXiv preprint arXiv:2304.10149\/} (2023).

\bibitem{liu2023summary}
{\sc Liu, Y., Han, T., Ma, S., Zhang, J., Yang, Y., Tian, J., He, H., Li, A., He, M., Liu, Z., et~al.}
\newblock Summary of chatgpt-related research and perspective towards the future of large language models.
\newblock {\em Meta-Radiology\/} (2023), 100017.

\bibitem{liu2024dr}
{\sc Liu, Z., Wu, L., He, M., Guan, Z., Zhao, H., and Feng, N.}
\newblock Dr. e bridges graphs with large language models through words.
\newblock {\em arXiv preprint arXiv:2406.15504\/} (2024).

\bibitem{ma2019jointly}
{\sc Ma, W., Zhang, M., Cao, Y., Jin, W., Wang, C., Liu, Y., Ma, S., and Ren, X.}
\newblock Jointly learning explainable rules for recommendation with knowledge graph.
\newblock In {\em The world wide web conference\/} (2019), pp.~1210--1221.

\bibitem{mcauley2013hidden}
{\sc McAuley, J., and Leskovec, J.}
\newblock Hidden factors and hidden topics: understanding rating dimensions with review text.
\newblock In {\em Proceedings of the 7th ACM conference on Recommender systems\/} (2013), pp.~165--172.

\bibitem{mcauley2015image}
{\sc McAuley, J., Targett, C., Shi, Q., and Van Den~Hengel, A.}
\newblock Image-based recommendations on styles and substitutes.
\newblock In {\em Proceedings of the 38th international ACM SIGIR conference on research and development in information retrieval\/} (2015), pp.~43--52.

\bibitem{peake2018explanation}
{\sc Peake, G., and Wang, J.}
\newblock Explanation mining: Post hoc interpretability of latent factor models for recommendation systems.
\newblock In {\em Proceedings of the 24th ACM SIGKDD International Conference on Knowledge Discovery \& Data Mining\/} (2018), pp.~2060--2069.

\bibitem{radford2019language}
{\sc Radford, A., Wu, J., Child, R., Luan, D., Amodei, D., Sutskever, I., et~al.}
\newblock Language models are unsupervised multitask learners.

\bibitem{reimers-2019-sentence-bert}
{\sc Reimers, N., and Gurevych, I.}
\newblock Sentence-bert: Sentence embeddings using siamese bert-networks.
\newblock In {\em Proceedings of the 2019 Conference on Empirical Methods in Natural Language Processing\/} (11 2019), Association for Computational Linguistics.

\bibitem{ren2017social}
{\sc Ren, Z., Liang, S., Li, P., Wang, S., and de~Rijke, M.}
\newblock Social collaborative viewpoint regression with explainable recommendations.
\newblock In {\em Proceedings of the tenth ACM international conference on web search and data mining\/} (2017), pp.~485--494.

\bibitem{shin2020autoprompt}
{\sc Shin, T., Razeghi, Y., Logan~IV, R.~L., Wallace, E., and Singh, S.}
\newblock Autoprompt: Eliciting knowledge from language models with automatically generated prompts.
\newblock {\em arXiv preprint arXiv:2010.15980\/} (2020).

\bibitem{singh2018posthoc}
{\sc Singh, J., and Anand, A.}
\newblock Posthoc interpretability of learning to rank models using secondary training data.
\newblock {\em arXiv preprint arXiv:1806.11330\/} (2018).

\bibitem{strubell2019energy}
{\sc Strubell, E., Ganesh, A., and McCallum, A.}
\newblock Energy and policy considerations for deep learning in nlp.
\newblock {\em arXiv preprint arXiv:1906.02243\/} (2019).

\bibitem{sun2019bert4rec}
{\sc Sun, F., Liu, J., Wu, J., Pei, C., Lin, X., Ou, W., and Jiang, P.}
\newblock Bert4rec: Sequential recommendation with bidirectional encoder representations from transformer.
\newblock In {\em Proceedings of the 28th ACM international conference on information and knowledge management\/} (2019), pp.~1441--1450.

\bibitem{tao2019fact}
{\sc Tao, Y., Jia, Y., Wang, N., and Wang, H.}
\newblock The fact: Taming latent factor models for explainability with factorization trees.
\newblock In {\em Proceedings of the 42nd international ACM SIGIR conference on research and development in information retrieval\/} (2019), pp.~295--304.

\bibitem{tintarev2015explaining}
{\sc Tintarev, N., and Masthoff, J.}
\newblock Explaining recommendations: Design and evaluation.
\newblock In {\em Recommender systems handbook}. Springer, 2015, pp.~353--382.

\bibitem{touvron2023llama}
{\sc Touvron, H., Lavril, T., Izacard, G., Martinet, X., Lachaux, M.-A., Lacroix, T., Rozi{\`e}re, B., Goyal, N., Hambro, E., Azhar, F., et~al.}
\newblock Llama: Open and efficient foundation language models.
\newblock {\em arXiv preprint arXiv:2302.13971\/} (2023).

\bibitem{vaswani2017attention}
{\sc Vaswani, A., Shazeer, N., Parmar, N., Uszkoreit, J., Jones, L., Gomez, A.~N., Kaiser, {\L}., and Polosukhin, I.}
\newblock Attention is all you need.
\newblock {\em Advances in neural information processing systems 30\/} (2017).

\bibitem{wang2018reinforcement}
{\sc Wang, X., Chen, Y., Yang, J., Wu, L., Wu, Z., and Xie, X.}
\newblock A reinforcement learning framework for explainable recommendation.
\newblock In {\em 2018 IEEE international conference on data mining (ICDM)\/} (2018), IEEE, pp.~587--596.

\bibitem{wang2018tem}
{\sc Wang, X., He, X., Feng, F., Nie, L., and Chua, T.-S.}
\newblock Tem: Tree-enhanced embedding model for explainable recommendation.
\newblock In {\em Proceedings of the 2018 world wide web conference\/} (2018), pp.~1543--1552.

\bibitem{wang2023recmind}
{\sc Wang, Y., Jiang, Z., Chen, Z., Yang, F., Zhou, Y., Cho, E., Fan, X., Huang, X., Lu, Y., and Yang, Y.}
\newblock Recmind: Large language model powered agent for recommendation.
\newblock {\em arXiv preprint arXiv:2308.14296\/} (2023).

\bibitem{wei2022chain}
{\sc Wei, J., Wang, X., Schuurmans, D., Bosma, M., Xia, F., Chi, E., Le, Q.~V., Zhou, D., et~al.}
\newblock Chain-of-thought prompting elicits reasoning in large language models.
\newblock {\em Advances in neural information processing systems 35\/} (2022), 24824--24837.

\bibitem{wei2024llmrec}
{\sc Wei, W., Ren, X., Tang, J., Wang, Q., Su, L., Cheng, S., Wang, J., Yin, D., and Huang, C.}
\newblock Llmrec: Large language models with graph augmentation for recommendation.
\newblock In {\em Proceedings of the 17th ACM International Conference on Web Search and Data Mining\/} (2024), pp.~806--815.

\bibitem{wu2024exploring}
{\sc Wu, L., Qiu, Z., Zheng, Z., Zhu, H., and Chen, E.}
\newblock Exploring large language model for graph data understanding in online job recommendations.
\newblock In {\em Proceedings of the AAAI Conference on Artificial Intelligence\/} (2024), vol.~38, pp.~9178--9186.

\bibitem{wu2023learning}
{\sc Wu, L., Zhao, H., Li, Z., Huang, Z., Liu, Q., and Chen, E.}
\newblock Learning the explainable semantic relations via unified graph topic-disentangled neural networks.
\newblock {\em ACM Transactions on Knowledge Discovery from Data 17}, 8 (2023), 1--23.

\bibitem{wu2023survey}
{\sc Wu, L., Zheng, Z., Qiu, Z., Wang, H., Gu, H., Shen, T., Qin, C., Zhu, C., Zhu, H., Liu, Q., et~al.}
\newblock A survey on large language models for recommendation.
\newblock {\em arXiv preprint arXiv:2305.19860\/} (2023).

\bibitem{xian2019reinforcement}
{\sc Xian, Y., Fu, Z., Muthukrishnan, S., De~Melo, G., and Zhang, Y.}
\newblock Reinforcement knowledge graph reasoning for explainable recommendation.
\newblock In {\em Proceedings of the 42nd international ACM SIGIR conference on research and development in information retrieval\/} (2019), pp.~285--294.

\bibitem{zhang2020explainable}
{\sc Zhang, Y., Chen, X., et~al.}
\newblock Explainable recommendation: A survey and new perspectives.
\newblock {\em Foundations and Trends{\textregistered} in Information Retrieval 14}, 1 (2020), 1--101.

\bibitem{zhang2014explicit}
{\sc Zhang, Y., Lai, G., Zhang, M., Zhang, Y., Liu, Y., and Ma, S.}
\newblock Explicit factor models for explainable recommendation based on phrase-level sentiment analysis.
\newblock In {\em Proceedings of the 37th international ACM SIGIR conference on Research \& development in information retrieval\/} (2014), pp.~83--92.

\end{thebibliography}

\newpage
\appendix
\section{Seven metrics and their description problems} \label{appendix:seven metrics}
The seven metrics and their complete description problems are shown in Table \ref{tab:metrics and problems}, where Model{x} represents the number name of each model, which is used to hide the model information.

\begin{table}[h]
\caption{metrics and description problems}
\label{tab:metrics and problems}
\centering
\setlength{\tabcolsep}{6pt} % Adjust column separation if necessary
\begin{tabular}{c|l}
\toprule
\textbf{Metric}  & \textbf{Description Problems} \\
\midrule
clarity          & Is the explanation provided by Model{x} easy to understand? \\
detail           & Is the explanation of Model{x} detailed enough? \\
effectiveness    & Does the explanation of Model{x} help you understand the reason for the recommendation? \\
relevance        & Is the explanation of Model{i} accurate and consistent? \\
logic            & Is the explanation of Model{i} reasonable in structure and logically rigorous? \\
trust            & Do you trust the recommendations provided by Model{x}? \\
satisfaction     & What is your overall satisfaction with Model{x}? \\
\bottomrule
\end{tabular}
\end{table}

\section{Sample Information and explanation of framework output} \label{appendix:case study}

\textbf{1. Sample Information}

\textbf{User's interaction sequence}: [`Condemned: Criminal Origins', `BioShock® 2', `Alan Wake', `Mortal Kombat Komplete Edition', `The Dig®', `World of Goo', `Mass Effect 2', `STAR WARS™ Jedi Knight - Mysteries of the Sith™', `STAR WARS™ Jedi Knight - Jedi Academy™', `Machinarium', `Samorost 2', `Thief', `Hector: Badge of Carnage - Full Series', `Antichamber', `Call of Duty®', `Assassin's Creed® III', `Murdered: Soul Suspect', `Home', `Crysis 2 - Maximum Edition', `Botanicula', `LISA', `Home is Where One Starts...', `Call of Duty®: Black Ops', `The Novelist', `STAR WARS™ - The Force Unleashed™ Ultimate Sith Edition', `Deus Ex: Human Revolution - Director's Cut', `Borderlands', `The Talos Principle', `To the Moon', `The Typing of The Dead: Overkill', `Electronic Super Joy', `BIT.TRIP RUNNER', `A Story About My Uncle', `Deadly Premonition: The Director's Cut', `Deadly Premonition: The Director's Cut', `Remember Me', `BIT.TRIP FLUX']

\textbf{User preferences}: \{`Action and Adventure Games': 0.3561, `Sci-Fi Themed Games': 0.3094, `Indie Games with Unique Storylines': 0.1041, `Puzzle and Platform Games': 0.108, `Games with Strong Narrative Elements': 0.1225\}

\textbf{Target item}: Far Cry® 2: Fortune's Edition

\textbf{Rank in candidate}: 5/101

\textbf{2. Explanation of framework output}

\textbf{Step 1}:

\textbf{Preference 1}: Action and Adventure Games

\textbf{Analysis}: The user has a strong inclination towards action and adventure games as seen from several titles in their historical interaction sequence, such as `Mortal Kombat Komplete Edition', `Call of Duty®', `Assassin's Creed® III', and `Crysis 2 - Maximum Edition'. These games offer thrilling experiences with dynamic gameplay and immersive worlds that capture the player's interest. The user likely enjoys the adrenaline rush and engagement provided by these genres.

\textbf{Preference 2}: Sci-Fi Themed Games

\textbf{Analysis}: The user has shown a solid preference for sci-fi themed games, as evidenced by titles like `BioShock® 2', `Mass Effect 2', `STAR WARS™ Jedi Knight - Jedi Academy™', and `Deus Ex: Human Revolution - Director's Cut'. These games typically feature futuristic settings, advanced technology, and complex plotlines, which suggest that the user finds enjoyment in speculative and imaginative scenarios.

\textbf{Preference 3}: Indie Games with Unique Storylines

\textbf{Analysis}: The user's interest in indie games with unique storylines is evident from their interactions with titles such as `Antichamber', `World of Goo', `Machinarium', and `To the Moon'. These games often feature distinct art styles and innovative narratives that mainstream games might not offer, indicating the user's appreciation for creativity and original storytelling.

\textbf{Preference 4}: Puzzle and Platform Games

\textbf{Analysis}: The user has shown affinity for puzzle and platform games through their interaction with titles like `The Talos Principle', `BIT.TRIP RUNNER', `World of Goo', and `A Story About My Uncle'. These games challenge the player's problem-solving skills and often feature platforming elements, which the user seems to enjoy.

\textbf{Preference 5}: Games with Strong Narrative Elements

\textbf{Analysis}: The user appears to have a penchant for games with strong narrative elements, as seen from `The Novelist', `Remember Me', `Murdered: Soul Suspect', and `The Dig®'. These games focus heavily on storytelling and character development, indicating that the user values a compelling story along with gameplay.

\textbf{Step 2}:

\textbf{Target item introduction}:

`Far Cry® 2: Fortune's Edition' is an open-world first-person shooter game set in a fictional African country embroiled in conflict. The game features a dynamic weather system, realistic fire propagation, and a mission-driven storyline.

\textbf{Preference Fitness}:

1. \textbf{Action and Adventure Games}: 0.9

\textbf{Reason}: `Far Cry® 2' is a quintessential action and adventure game with open-world exploration, combat, and various missions.

2. \textbf{Sci-Fi Themed Games}: 0.2

\textbf{Reason}: `Far Cry® 2' is not set in a sci-fi universe and lacks futuristic elements, reducing its appeal based on this preference.

3. \textbf{Indie Games with Unique Storylines}: 0.3

\textbf{Reason}: While `Far Cry® 2' has a strong narrative, it is not an indie game, nor does it possess a particularly unique storyline compared to the user's indie game preferences.

4. \textbf{Puzzle and Platform Games}: 0.1

\textbf{Reason}: The game does not focus on puzzles or platforming mechanics, making it a poor fit for this preference.

5. \textbf{Games with Strong Narrative Elements}: 0.7

\textbf{Reason:} `Far Cry® 2' does have a compelling narrative involving mercenaries, war, and survival, providing some level of satisfaction for this preference.

\textbf{Step 3}:

\textbf{Interaction probability}: Medium

\textbf{Reason}: Given the strong alignment with the user's top preferences (Action and Adventure Games with a weight of 0.3561 and a high fitness score of 0.9, and Games with Strong Narrative Elements with a weight of 0.1225 and a fitness score of 0.7), there is a significant probability that the user would interact with `Far Cry® 2: Fortune's Edition'. The lower fitness scores for Sci-Fi Themed Games, Indie Games with Unique Storylines, and Puzzle and Platform Games reduce overall interaction probability but do not overpower the strong inclinations.

\textbf{Step 4}:

\textbf{Recommendation}: Based on your strong preference for action-packed adventure games and those with compelling narratives, `Far Cry® 2: Fortune's Edition' aligns well with your gaming tastes. While it may lack the sci-fi elements and unique indie storytelling you're also fond of, the game's immersive open-world experience and engaging missions make it a solid option for your next adventure. Enjoy the exploration and tactical combat in the realistic setting of Far Cry® 2!

\end{document}